\documentclass[sigconf,nonacm,pbalance]{acmart}

\usepackage[justification=centering]{caption}
\usepackage{subcaption}
\usepackage[frozencache]{minted}
\usepackage{multirow}
\usepackage{pifont}%
\usepackage{svg}
\usepackage{tikz}
\usetikzlibrary{matrix}
\newcommand{\cmark}{\ding{51}}%
\newcommand{\xmark}{\ding{55}}%

\begin{document}

\title{Blaze: Compiling JSON Schema for 10× Faster Validation}

\author{Juan Cruz Viotti}
\email{jviotti@sourcemeta.com}
\orcid{0000-0003-1732-0222}
\affiliation{%
  \institution{Sourcemeta Ltd}
  \streetaddress{71-75 Shelton Street}
  \city{London}
  \country{UK}
  \postcode{WC2H 9JQ}
}
\author{Michael J. Mior}
\email{mmior@mail.rit.edu}
\orcid{0000-0002-4057-8726}
\affiliation{%
  \institution{Rochester Institute of Technology}
  \streetaddress{102 Lomb Memorial Drive}
  \city{Rochester}
  \state{New York}
  \country{USA}
  \postcode{14612-5608}
}

\renewcommand{\shortauthors}{Viotti et al.}

\begin{abstract}
JSON Schemas provide useful guardrails for developers of Web APIs to guarantee that the semi-structured JSON input provided by clients matches a predefined structure.
This is important both to ensure the correctness of the data received as input and also to avoid potential security issues from processing input that is not correctly validated.
However, this validation process can be time-consuming and adds overhead to every request.
Different keywords in the JSON Schema specification have complex interactions that may increase validation time.
Since popular APIs may process thousands of requests per second and schemas change infrequently, we observe that we can resolve some of the complexity ahead of time in order to achieve faster validation.

Our JSON Schema validator, Blaze, compiles complex schemas to an efficient representation in seconds to minutes, adding minimal overhead at build time.
Blaze incorporates several unique optimizations to reduce the validation time by an average of approximately 10$\times$ compared existing validators on a variety of datasets.
In some cases, Blaze achieves a reduction in validation time of multiple orders of magnitude compared to the next fastest validator.
We also demonstrate that several popular validators produce incorrect results in some cases, while Blaze maintains strict adherence to the JSON Schema specification.
\end{abstract}

\maketitle

\section{Introduction}\label{sec:intro}

Web APIs commonly accept payloads in the semi-structured JSON format that enable significant flexibility in the input that is accepted~\cite{song2023restgpt,RESTAPIs,TestingREST}.
This is achieved with nested objects and arrays in JSON data structures that do not require the exact format to be clearly defined.
Despite this flexibility, validating that the data received by an API is in a specific format is important for both correctness and security.
In order to validate payloads, such APIs commonly make use of the JSON Schema standard.
JSON Schema allows the API developer to validate the structure of a JSON payload including required properties, data types, and other features.
The declarative approach of JSON Schema makes such validation easier to write and maintain than an explicit set of validation instructions.
JSON Schemas are expressed in JSON format with a set of keywords that restrict the set of JSON documents that are valid according to the schema.
An example of a simple JSON Schema with both valid and invalid documents is shown in Figure~\ref{fig:json_schema_example}.
This particular example is very straightforward, but interpreting schemas at runtime can be complex due to the interaction of various keywords in the JSON Schema specification.

There are a wide variety of tools and standards that make use of JSON Schema.
For example, OpenAPI~\footnote{\url{https://www.openapis.org}} is a commonly used format to define the structure of Web APIs.
OpenAPI uses JSON Schema to describe request and response payloads.
Validating requests according to JSON Schemas is a common practice, placing JSON Schema validation in the critical path of responding to a request.
This means that any delay introduced by validation has an impact on latency.
Numerous studies have shown that even small increases in latency can have a negative impact on the performance perceived by users~\cite{arapakis2014impact,nah2004study}.
Our goal with Blaze is to minimize the latency of validating documents, even for large, complex schemas.

Attouche et al.~\cite{modernjson} have shown that the latest dialect (2020-12) of the JSON Schema specification is PSPACE-complete with respect to the size of the schema.
In particular, this is caused by the use of dynamic references, which are used to implement generic types or extend recursive schemas.
However, they also identify that some validators incur a significant overhead from implementing dynamic references even when a particular schema does not actually make use of this feature.
As we discuss further throughout the paper, we use open source schemas collected from GitHub to analyze the usage of JSON Schema in practice.
Among the more than 31,000 schemas collected, we found only 10 instances of dynamic references.
This means that some validators incur significant overhead for a feature of JSON Schema that is rarely used.
We do not focus specifically on dynamic references in this work, but we do describe a method for converting some instances of dynamic references to static references during compilation.
Our approach for doing so incurs no overhead at validation time.
We also implement several other optimizations ahead of time through a precompilation process that transforms a provided schema into a list of instructions that enable efficient validation.

\begin{figure}
  \centering
  \begin{subfigure}[b]{0.4\textwidth}
      \centering
      \begin{minted}[fontsize=\scriptsize]{json}
          {
            "$schema": "https://json-schema.org/draft/2020-12/schema",
            "properties": {
              "firstName": {"type": "string", "maxLength": 100},
              "middleName": {"type": "string"},
              "lastName": {"type": "string", "maxLength": 100},
              "age": {"type": "integer", "minimum": 0}
            },
            "required": ["firstName", "lastName"]
          }
      \end{minted} 
      \caption{JSON Schema example}
      \Description[JSON Schema]{A JSON Schema defining firstName, middleName, lastName, and age properties. Only firstName and lastName are required}
  \end{subfigure}
  \\~\\
  \begin{subfigure}[b]{0.2\textwidth}
      \centering
      \begin{minted}[fontsize=\scriptsize]{json}
          {
            "firstName": "Douglas",
            "lastName": "Crockford",
            "age": 69
          }
      \end{minted}
      \caption{Valid document}
      \Description[A valid JSON document]{A document containing firstName and lastName valid according to the previous schema}
  \end{subfigure}
  \begin{subfigure}[b]{0.2\textwidth}
      \centering
      \begin{minted}[fontsize=\scriptsize]{json}
          {
            "firstName": "Jason",
            // Missing lastName
            "age": 20
          }
      \end{minted}
      \caption{Invalid document}
      \Description[An valid JSON document]{A document containing only firstName and age that is valid according to the previous schema}
  \end{subfigure}
  \caption{}\label{fig:json_schema_example}
\end{figure}

While validation happens often, schemas change relatively infrequently so we can afford to invest extra time in the compilation process in order to achieve faster validation in the future.
Across the schemas we analyzed from GitHub, the average time between commits to an individual schema over 65 days.
Since schemas are changed infrequently, we can safely invest time in schema compilation to use for later validation.
This is the approach that we take with Blaze.
Our contributions in this paper are as follows:
1) an optimized low-level DSL for schema validation designed for semi-structured data validation,
2) a mapping from JSON Schema to our DSL that incorporates several unique optimizations,
and 3) an execution engine that can efficiently validate documents according to instructions from our DSL.

Section~\ref{sec:language} describes the schema validation language that is the result of schema compilation.
We then describe how we compile from JSON Schema to this language in Section~\ref{sec:compilation}.
Section~\ref{sec:optimization} introduces several important optimizations to our compilation process and Section~\ref{sec:execution} describes the Blaze executor along with a complete example.
Section~\ref{sec:evaluation} provides a comprehensive evaluation comparing Blaze to many existing evaluators using a variety of schemas.
Finally, we discuss related and future work and conclude in Sections~\ref{sec:related}-\ref{sec:conclusion}.

\section{Schema Validation Language}\label{sec:language}

Like JSON Schema, by default, the DSL we define for validation is a constraint language~\cite{constraintlanguage}.
In other words, the language is permissive, meaning that any value is acceptable.
To restrict the set of documents accepted, we introduce a list of instructions that potentially cause validation to fail if the document does not meet a specific set of conditions.
Each instruction contains a location in the document it applies to and then some assertion on the contents of that portion of the document.
In the following sections, we first introduce basic instructions that apply to single values and then supplement these with instructions that loop over nested values and logically combine the results from multiple instructions.
Note that by convention all our instructions start with uppercase letters while JSON Schema keywords start with lowercase letters.
We also introduce some basic optimizations we apply to these instructions to improve performance.
Further higher-level optimizations are discussed in Section~\ref{sec:optimization}.

\subsection{Basic Instructions}\label{subsec:basic_inst}

The most basic instructions implemented in our validation language operate on single values.
For example, the \texttt{TypeAny} instruction validates that a value is one of a given set of types.
The instruction \texttt{TypeAny /foo ["string", "number"]} validates that a value of the property \texttt{"foo"} is either a string or a number.
Similarly, \texttt{EqualsAny} validates that a value is one of a given set of specific values.
The other instructions apply only to a specific type of value and are ignored for values of other types.
These instructions have preconditions that check whether a value has a particular type before continuing execution.
As we later show in Section~\ref{sec:compilation}, this will allow us to support JSON Schema keywords that only apply to specific types of values.
The remaining basic instructions are summarized in Table~\ref{tab:basic}.
Each of these instructions have a precondition on their corresponding type.

\begin{table*}[h]
{\small
    \centering
    \begin{tabular}{|c|c|c|}
        \hline
        Instruction & Type & Purpose \\ \hline\hline
        \texttt{DefinesAny} & object & specific properties exist \\ \hline
        \texttt{PropertyDependencies} & object & if a property exist, other properties must also exist \\ \hline
        \texttt{ObjectSize} & object & validates the number of properties in an object \\ \hline
        \texttt{PropertyType} & object & an object has a property of a specific type \\ \hline
        \texttt{Regex} & string & a specific regular expression matches \\ \hline
        \texttt{StringSize} & string & validates the length of a string \\ \hline
        \texttt{StringType} & string & validates complex string formats such as URIs \\ \hline
        \texttt{Unique} & array & all array elements are unique \\ \hline
        \texttt{ArraySize} & array & validates the size of an array \\ \hline
        \texttt{Less}/\texttt{Greater}/\texttt{Equal} & number & validates the range of numeric values \\ \hline
        \texttt{Divisible} & number & validates whether a number is divisible by a given value \\ \hline
    \end{tabular}
    }
    \caption{Basic instructions and their corresponding types}\label{tab:basic}
\end{table*}

\subsection{Loops}\label{subsec:loops}

In several cases, we do not know in advance the set of values that must be evaluated to validate a schema.
This occurs with arrays or objects where the set of keys is not predefined by the schema in advance.
There are three specific scenarios we need to consider.
Looping over object keys, object values, and array items.
Looping over the keys of an object (ignoring values) is implemented using the \texttt{LoopKeys} instruction.
Each key is validated against a given set of string instructions (e.g., the key must match a given regular expression or meet minimum/maximum length requirements).
There are four different variants of looping over values of an object.
\texttt{LoopProperties} validates \emph{all} values according to a single set of instructions.
\texttt{Loop\-Properties\-Except} validates all values except those whose corresponding keys matching either a given regular expression or a static list of keys.
\texttt{Loop\-Properties\-Regex} validates only values whose corresponding keys match a given regular expression.
These first three variants loop over the schema, with each instruction matched against the instance being validated.
The final two variants loop over the instance and then look up instructions to execute within the schema.
\texttt{Loop\-Properties\-Match} that contains a set of keys and a list of instructions for corresponding values to validate against.
The final loop instruction is a variant, \texttt{Loop\-Properties\-Match\-Closed} that is used when all properties in an object must be explicitly defined in advance and the loop must validate all properties.

For arrays, there are two types of loop instructions.
The first is \texttt{LoopItems} which validates all items in an array against a given set of instructions.
A variant of this instruction, \texttt{LoopItemsFrom}, skips the first $n$ items in the array, which as we show later, is useful for validating against some JSON Schema keywords.

\subsection{Logical Operators}\label{subsec:logical}

In some cases, it is desirable to combine instructions based on logical operators, such as requiring any of a given set of conditions to hold.
Accordingly, we define a set of instructions that conditionally execute other instructions based on a set of conditions.
Firstly, we have the \texttt{Logical\-Condition} instruction that executes a different set of instructions based on whether a given condition is true or false.
Second, we have a set of instructions to combine the results of multiple steps using logical operators.
These include \texttt{LogicalAnd}, \texttt{LogicalOr}, \texttt{LogicalXor}, and \texttt{Logical\-Not}.
Note that these operators will short circuit where possible if failure to validate is guaranteed without evaluating all steps.
For example, if the first instruction evaluated inside \texttt{LogicalAnd} evaluates to false, execute does not need to continue since the final result can already be determined.

\subsection{Control Flow}\label{subsec:control}

It is common when validating semi-structured data to reuse part of a schema to validate similar structures in different locations with a document.
Accordingly, we introduce the \texttt{ControlLabel} and \texttt{ControlJump} to enable this use case.
The \texttt{ControlLabel} instruction contains a list of child instructions that are executed when the label is first encountered.
The \texttt{ControlJump} instruction is also able to return these labeled instructions at any future point when the schema is evaluated.
As we discuss in Section~\ref{sec:compilation}, we use these instructions to support references within a JSON Schema.

\subsection{Instruction Set Optimization}

\begin{table}[h]
{\small
    \centering
    \begin{tabular}{|c|c|}
        \hline
        Instruction & Condition \\ \hline\hline
        \texttt{WhenType} & Specific type \\ \hline
        \texttt{WhenDefines} & Object that defines specific properties \\ \hline
        \texttt{WhenArraySizeGreater} & Array with a minimum size \\ \hline
        \texttt{WhenArraySizeEqual} & Array with a maximum size \\ \hline
    \end{tabular}
    }
    \caption{Logical condition optimizations}\label{tab:condition_opt}
\end{table}

In general, we found that executing a smaller number of slightly more complex instructions leads to more efficient validation.
This is largely because dynamic dispatch of individual instructions during validation can be costly.
Accordingly, we introduce variants of several of our instructions analogous to the CISC approach of many modern processors~\cite{george1990overview}.
For example, we define an instruction \texttt{StringBounds} that combines several separate instructions: \texttt{Type} to validate that the value is a string along with \texttt{StringSizeGreater} and \texttt{StringSizeLess} to validate that the length of the string is within a given range.
In this vein, we also define variants of instructions ending with \texttt{Any} when only a single value is required.
While the \texttt{EqualsAny} instruction checks if a value is one of a given list of values, the \texttt{Equals} instruction checks against a single value.
This avoids a loop and yields modest efficiency gains.
Another scenario we optimize is \texttt{LogicalCondition}.
There are a number of specific conditions for which we define separate instructions as is shown in Table~\ref{tab:condition_opt}.
We plan to explore further opportunities for optimization in the future.

\section{JSON Schema Compilation}\label{sec:compilation}

Our approach to convert JSON Schemas to our validation language follows a similar approach to the formalization of Attouche et al.\ for the 2020-12 dialect~\cite{modernjson} of JSON Schema.
Unless otherwise specified, all schemas we refer to use this most recent dialect.
Blaze also supports JSON Schema dialects 4, 6, 7, and 2019-09 in a similar fashion to 2020-12 which we discuss here.
In this work, we focus only on determining whether a document is valid and we ignore annotations and dynamic references.
While Blaze supports both of these features, we save their discussion for future work.
The formalization of Attouche et al.\ allows individual keywords to be evaluated sequentially subject to some ordering constraints~\cite{staticanalysis}.
We start with keywords that can be evaluated independent from others and then describe the operation of dependent keywords and how we handle references within documents.

\subsection{Independent Keywords}

Some of the keywords in JSON Schema can be evaluated against a value in any order without consideration for the effect of adjacent keywords.
These keywords are referred to as \emph{independent} keywords.
We can exploit this independence to place instructions for more costly keywords first, saving processing time in case an earlier keyword causes validation to fail.
For example, checking string length before validating a regular expression.
There are two types of independent keywords.
The first are \emph{assertions} that have atomic values and define simple constraints.
The second are applicators which themselves contain schemas to validate more complex values.

\subsubsection{Assertions}\label{subsubsec:assertions}

For an example of an assertion, consider \texttt{minimum} that requires a numeric value be greater than a lower bound.
In JSON Schema, many assertion only apply to a value of a given type.
In this case, \texttt{minimum} only applies to numeric values.
We therefore convert the application of the \texttt{minimum} keyword to the keyword \texttt{Greater} in our validation language, that also only applies to numeric values.
Each assertion is mapped to instructions in our validation language that validate the assertion.
Note that in the process of converting from assertions to instructions in our validation language, we also apply several static optimizations.
For example, if the schema defines the type of a value to be anything other than \texttt{integer} or \texttt{number}, the \texttt{minimum} assertion is redundant can be ignored and no validation instructions generated since it only applies to numeric values.
Such unnecessary assertions sometimes occur due to errors in authoring the schema.
We plan to make use of these observations to develop a tool to highlight possible errors or optimizations in a schema.

\subsubsection{Applicators}

Several applicators are also independent keywords, but the value of an applicator may be a schema.
For example, the \texttt{propertyNames} keyword is an applicator that applies a schema to the keys of a JSON object. 
When compiling applicators, we recursively compile the corresponding subschema.
Note that applicators may be nested recursively and our compilation process continues recursively as needed.

The applicators \texttt{anyOf}, \texttt{oneOf}, \texttt{allOf}, and \texttt{not} compile to the logical instructions \texttt{LogicalOr}, \texttt{LogicalXor}, \texttt{LogicalAnd}, and \texttt{Logical\-Not} respectively.
These applicators indicate that a location in the JSON document specify which of a number of schemas a given location in a document must match (or not).
As mentioned previously, these instructions can short-circuit evaluation and skip instructions that are not necessary to determine whether a document is valid.

For conditional application of schemas based on \texttt{if}, \texttt{then}, and \texttt{else}, we handle all compilation upon encountering the \texttt{if} applicator.
The \texttt{if} applicator specifies a schema.
If the value at the instance location is valid according to the schema, then the schema specified under \texttt{then} is applied, otherwise the schema specified under \texttt{else} is applied.
To compile into our validation language, we recursively compile each of these schemas and then generate a \texttt{LogicalCondition} instruction that jumps to the appropriate schema for further validation.
As a minor optimization, if there is no schema specified for \texttt{if}, then it will be considered true, and we can avoid compiling the \texttt{else} case.

For arrays, \texttt{contains} is an applicator that specifies a schema that must validate against at least one array element.
The keywords \texttt{minContains} and \texttt{maxContains} can be used to control the number of required elements.
We emit the \texttt{LoopContains} instruction to validate that the array contains the required number of elements.
We note that several small optimizations are possible here.
First, if \texttt{minContains} is 0 and \texttt{maxContains} is not set, we do not need to emit any instructions at all.
Second, if the schema provided to \texttt{contains} is \texttt{true} (that is, all values are accepted), then we only need to validate the array length and \texttt{minContains} and \texttt{maxContains} have the same effect as \texttt{minItems} and \texttt{maxItems}.
The other independent array applicator is \texttt{prefixItems} that validates that items in a prefix of an array match a given array of schemas.
In this case, we recursively compile each schema and then generate an \texttt{ArrayPrefix} instruction that performs the validation of all items.

Objects also have several independent applicators that can be evaluated in any order.
The \texttt{properties} and \texttt{pattern\-Properties} applicators are very similar.
Both specify schemas that must be valid against values inside an object.
The \texttt{properties} applicator specifies string literals for keys while \texttt{pattern\-Properties} specifies regular expressions to match keys.
Note that if an expression listed under the \texttt{properties} keyword also matches a regular expression listed under \texttt{pattern\-Properties}, then the value must be valid according to both schemas.
This is what allows these two keywords to be evaluated independently.
The schemas for each value in the \texttt{properties} and \texttt{pattern\-Properties} applicators are recursively compiled and the \texttt{LoopProperties} and \texttt{Loop\-Properties\-Except} instructions are used for their evaluation.

Finally, the \texttt{propertyNames} keyword specifies a schema that must be validated.
In this case, we recursively compile this schema and generate a \texttt{LoopKeys} instruction to validate all keys in an object against the schema.

\subsection{Dependent Keywords}

There are two different types of dependent keywords: first-level, and second-level dependent.
Along with independent keywords, these define three tiers of keywords we must consider.
Independent keywords can be evaluated in any order without considering any other keywords.
First-level dependent keywords have some dependencies on other keywords.
However, we introduce mechanisms in our compilation process to allow instructions compiled from these keywords to also be executed in any order.
Second-level dependent keywords may depend on the evaluation of the remainder of the schema and cannot necessary be resolved statically.

\subsubsection{First-Level Dependent Keywords}

There are two first-level dependent keywords: \texttt{additionalProperties} (for objects) and \texttt{items} (for arrays).
Both serve to validate values in objects (or arrays) that are not validated by other keywords.
The \texttt{additional\-Properties} applicator specifies a schema that every property not already validated by either \texttt{properties} or \texttt{patternProperties} must match.
While the behavior of \texttt{additionalProperties} depends on the values of the \texttt{prop\-erties} and \texttt{patternProperties} keywords, we can resolve this dependency statically to also allow instructions generated from \texttt{additional\-Properties} to be executed in any order.
To do so, we examine keywords adjacent to \texttt{additionalProperties} and collect the static set of keys and the regular expressions that are used in the \texttt{properties} and \texttt{pattern\-Properties} keywords.
We then generate an instruction that skips these properties to validate the remaining additional properties.

However, we also optimize the common case where the value of \texttt{additionalProperties} is a Boolean.
A value of \texttt{true} indicates that objects are permitted to have any additional properties without adhering to any particular schema.
In this case, we generate no instructions for validation.
When \texttt{additionalProperties} is set to \texttt{false}, only properties explicitly defined using the \texttt{properties} or \texttt{patternProperties} keywords are allowed.
In this case, we modify the instructions generated for these keywords to fail if any additional properties are encountered.

The \texttt{items} applicator functions similarly for arrays.
It validates all items in the array not already validated by \texttt{prefixItems}.
In this case, we recursively compile the schema for the items and use the \texttt{LoopItems} instruction to validate the items.
The first $n$ items are skipped by using the \texttt{LoopItemsFrom} instruction if the \texttt{prefixItems} applicator is also used with $n$ items.
In this way, the behavior of \texttt{items} is dependent on \texttt{prefixItems}, but the two can still be evaluated in any order.

\subsubsection{Second-Level Dependent Keywords}

The two second-level dependent applicators \texttt{unevaluatedProperties} and \texttt{unevaluated\-Items} serve a similar purpose.
They provide a schema that values in arrays or objects must be validated against if they have not already been validated by another keyword elsewhere in the schema.
This is similar to the first-level dependent keywords except that second-level dependent keywords can ``see through'' other applicators.
Consider the example use the of \texttt{unevaluatedProperties} keyword in Figure~\ref{fig:unevaluatedProperties} adapted from the JSON Schema documentation\footnote{\url{https://json-schema.org/understanding-json-schema/reference/object}}.
The properties \texttt{city} and \texttt{state} are contained inside the applicator \texttt{allOf}.
If the schema used the similar keyword \texttt{additional\-Properties} instead, the corresponding document would be invalid.
This is because the keyword \texttt{additional\-Properties} only applies to adjacent keywords such as \texttt{properties}.
It would not allow the properties that are defined inside the \texttt{allOf} applicator.

In contrast, the behavior of \texttt{unevaluatedProperties} is defined in terms of whether a property has been evaluated by another keyword, specifically, the keywords \texttt{properties}, \texttt{pattern\-Properties}, and \texttt{additional\-Properties}.
Most implementations take the approach of adding an \emph{annotation} to an object key when it has been evaluated by one of these keywords, regardless of whether that evaluation occurred inside another applicator.
Any key that does not have such an annotation applied is then evaluated by \texttt{unevaluated\-Properties}.
However, we find that maintaining annotations can add significant overhead and that in many cases, they are not necessary.
Instead, we can make a static pass over the schema to identify properties that will be evaluated by other keywords.
In the case of the schema in Figure~\ref{fig:unevaluatedProperties}, we can generate instructions for \texttt{unevaluatedProperties} in the same way as for \texttt{additional\-Properties} by statically identifying keys that are guaranteed to be evaluated.
In the case where \texttt{unevaluatedProperties} is \texttt{true} or all properties are guaranteed to be evaluated by other keywords, we do not generate any instructions for \texttt{unevaluatedProperties}.

\begin{figure}
  \centering
  \begin{subfigure}[t]{0.2\textwidth}
      \centering
      \begin{minted}[fontsize=\scriptsize]{json}
{"allOf": [{
    "type": "object",
    "properties": {
      "city": {"type": "string"},
      "state": {"type": "string"}
    }
  }],
  "properties": {
    "name": {"type": "string"}
  },
  "unevaluatedProperties": false}
      \end{minted} 
      \caption{JSON Schema}
      \Description[A valid JSON document]{A document valid according to the previous schema}
  \end{subfigure}
  \begin{subfigure}[t]{0.2\textwidth}
      \centering
      \begin{minted}[fontsize=\scriptsize]{json}
{
  "name": "Bob",
  "city": "Washington",
  "state": "DC"
}     
      \end{minted}
      \caption{Valid document}
      \Description[A valid JSON document]{A document valid according to the previous schema}
  \end{subfigure}
  \caption{Example use of \texttt{unevaluatedProperties}}\label{fig:unevaluatedProperties}
  \Description[unevaluatedProperties example]{A JSON schema and sample document showing the use of the unevaluatedProperties keyword}
\end{figure}

\subsection{Static References}\label{subsec:static_ref}

When references are used in a schema, it is advantageous to allow the reuse of validation instructions.
In fact, this is necessary for recursive references to avoid infinite loops in the generation of instructions.
To handle references, we use the instructions \texttt{ControlLabel} and \texttt{ControlJump}.
\texttt{ControlLabel} is used the first time the \texttt{\$ref} keyword is encountered.
The destination of the reference is recursively compiled into a set of instructions.
Any subsequent uses of \texttt{\$ref} throughout the schema use the \texttt{ControlJump} instruction to reuse the existing set of instructions.
However, for non-recursive references, we note that jumping between instructions can add additional overhead by reducing cache efficiency.
To mitigate this overhead, we eliminate the use of labels and jumps for non-recursive references that are repeated five or fewer times.
In these cases, we simply repeat the instructions compiled from the destination of the reference.
This avoids producing a significantly larger number of instructions for large numbers of repeated references.
We plan to explore improvements to this heuristic in future work.

\subsection{Dynamic References}

As past work has identified~\cite{defects}, dynamic references introduce significant challenges into JSON Schema validation.
Dynamic references are designed to allow targets of references to be changed when schemas are extended.
This means that the target of a reference cannot easily be statically determined at schema compilation time.
Instead, the resolution of the reference is dependent on the context determined at evaluation time.
This complicates validation process and it has been shown that validation with dynamic references is PSPACE-complete~\cite{modernjson,defects}.

We leave a full explanation and evaluation of our approach to handling dynamic references to future work.
However, we make two important observations.
First, as others have observed~\cite{modernjson}, it is possible to remove dynamic references from a schema.
In the general case, this can result in an exponential increase in the size of the schema.
Our second observation, as we discussed in Section~\ref{sec:intro}, is that dynamic references are very rarely used.
Furthermore, when dynamic references are used, there are often very few possible contexts for each dynamic reference.
In this work, we focus our evaluation on dynamic references with a single possible context.
In this case, a dynamic reference can be directly replaced by a static reference since we are guaranteed that the single available evaluation context will be used.
The dataset used in our evaluation in Section~\ref{sec:evaluation} contains two schemas, \texttt{openapi} and \texttt{cql2} that contain a dynamic reference with a single possible evaluation context that is transformed using this approach.

\subsection{Correctness}

As discussed previously, we base our formalization on that of Attouche et al.~\cite{modernjson}.
The key points addressed in their formulation are the order of keyword validation and the semantics of each individual keyword.
We started by creating careful mappings from each JSON Schema keyword into instructions that validate the keyword.
We carefully implement these instructions to ensure they follow the rules defined by the JSON Schema specification
As described in Section~\ref{sec:compilation}, we are able to evaluate instructions for independent and first-level dependent keywords in any order.
We guarantee that second-level dependent keywords are evaluated after their dependencies in order to ensure correct validation.
As discussed in Section~\ref{subsec:eval_correct}, we also validate our implementation against the official JSON Schema Test Suite~\cite{TestSuite}.
The test suite contains hundreds of tests for each JSON Schema version specifically designed to test difficult edge cases.
In particular, the test suite for the latest dialect supported by Blaze contains more than 1,200 tests.
As we discuss later, even several popular validators fail at least one case in this test suite.
Blaze passes every test case, increasing our confidence in its correctness.
We leave the possibility of formal verification of our implementation as future work.

\section{Optimizations}\label{sec:optimization}

We implemented Blaze in approximately 11,000 lines of C++20.
Our implementation is open source and available on GitHub~\footnote{\url{https://github.com/sourcemeta/blaze}} under the AGPL-3.0 license.
Although our description of Blaze has focused on the most current dialect (2020-12), we also support all previous dialects currently in use (4, 6, 7, and 2019-09).
We have put significant effort into optimizing each of the instructions executed by our validator.
While several optimizations are common to general software development, several unique optimizations stem from observations about the nature of data present in both JSON Schemas and typical JSON documents.
We describe these optimizations in the following subsections.
Specifically, we optimize the hash function used in data structures in our validator, unrolling instructions containing loops, and optimizing specific patterns found in regular expressions.

\subsection{Semi-perfect Hashing}\label{subsec:perfect_hashing}

One observation about both schemas and documents is that in most cases, the strings used as JSON keys are relatively short.
In our corpus of schemas we use for evaluation, 95\% of the keys defined in the schema are 13 characters or shorter.
We make use of this fact to optimize hashing of these strings for comparison and define a hash function that results in no collisions for short strings.
Unlike minimal perfect hash functions~\cite{fox1992practical}, we do not aim to avoid collisions in all cases, but focus on the most common case in practice.

The need to compare strings occurs frequently in schema validation for tasks such as checking the presence of required properties, whether a string matches a desired constant, and several other cases.
Hash functions used in library implementations such as the default MurmurHash\footnote{\url{https://github.com/aappleby/smhasher}} in the C++ standard library, are designed for general use and are not optimized for any particular use case.
We decide to optimize our hash functions for short strings at the expense of increased likelihood of collision for longer strings.
In fact, our goal is to avoid the need to compare strings at all in the common case by making our hash function one-to-one for small strings.

We start by defining the output size of our hash function to be 256 bits.
This is represented as four 64-bit bit integers (or two 128-bit integers on supported platforms), giving a total of 32 bytes.
We use the final 31 bytes when hashing strings that are 31 bytes or less.
In this case, the first byte is set to zero and the remaining bytes are copied directly from the string value.
This first byte is used in the case where the string is larger than 31 bytes.
In this case, the hash function is the sum of the size of the string and the first and last characters, modulo 255, plus one (ensuring the value is non-zero).
This allows computing a hash in constant time for longer strings.

To compare hashed strings, we first check that the first byte of each hash is zero, indicating that the two hashes both correspond to strings of 31 bytes or less.
At this point, we can compare the hash values directly to determine equality since the remaining bytes of the hash are exactly the bytes of strings.
Furthermore, we can do this comparison with basic integer comparisons instead of the need to check character by character.
If both strings are longer than 31 bytes, then it is still necessary to compare strings in the case of a hash collision.
Since this hash function is efficient to evaluate, we store the hash of strings as part of the process of parsing documents.
We then make use of this hash function and the optimized comparison anywhere strings are compared.
We also note that since documents generally have a small number keys, we make use of a vector data structure to store keys instead of a hash map since looping over the small number of entries is more efficient than dealing with the indirection inherent in hash tables.

While we would expect using a one byte hash for longer strings would increase the rate of collisions, this case is rare given our analysis of the common size of strings.
In the corpus of thousands of schemas we analyzed from GitHub, we found that over 98\% of keys defined JSON Schemas were less 32 characters.
This approach also has the advantage that we can compare short strings (less than 32 bytes) by comparing only their hash values.
We also note that the majority of JSON objects have a very small number of properties, so the rate of collisions is likely to remain low in practice.
We consider a common use case of our hash function which is to build a hash table for properties in object.
We looked at all JSON objects with more than one key across all of our test schemas and found that our hash function has a collision rate of less than 0.9\%.
MurmurHash has zero collisions on the same dataset.
However, our hash function still achieves better overall performance since we can compute the hash value in constant time.
Furthermore, when comparing short strings (less than 32 bytes), we only need to check their hash values and we can avoid string comparison entirely.
This is because when using our hash function, any strings less than 32 bytes are guaranteed to be equal if their hash values are equal.

\begin{figure}
\newcommand{\grayz}{\color{gray}0}
{\scriptsize\texttt{foobar}} \\
\begin{tikzpicture} [nodes={text width=0.4cm, text height=0.15cm, align=center, text depth=0.1cm},
         row sep=-\pgflinewidth, column sep=-\pgflinewidth](M)
         {\scriptsize
   \matrix (hash)[matrix of nodes, nodes={draw, anchor=center}]
         {
       \grayz & 102 & 111 & 111 & 98 & 97 & 114 & \grayz & \grayz & \grayz & \grayz & $\cdots$ & \grayz \\
    };
    }
\end{tikzpicture}
\\
{\scriptsize\texttt{corge}} \\
\begin{tikzpicture} [nodes={text width=0.4cm, text height=0.15cm, align=center, text depth=0.1cm},
         row sep=-\pgflinewidth, column sep=-\pgflinewidth]
         {\scriptsize
   \matrix (hash)[matrix of nodes, nodes={draw, anchor=center}]
         {
       \grayz & 99 & 111 & 114 & 103 & 101 & \grayz & \grayz & \grayz & \grayz & \grayz & $\cdots$ & \grayz \\
    };
    }
\end{tikzpicture}
{\scriptsize\texttt{thisisaverylongstringthatisgreaterthan31bytes}} \\
\begin{tikzpicture} [nodes={text width=0.4cm, text height=0.15cm, align=center, text depth=0.1cm},
         row sep=-\pgflinewidth, column sep=-\pgflinewidth]
         {\scriptsize
   \matrix (hash)[matrix of nodes, nodes={draw, anchor=center}]
         {
       20 & \grayz & \grayz & \grayz & \grayz & \grayz & \grayz & \grayz & \grayz & \grayz & \grayz & $\cdots$ & \grayz \\
    };
    }
\end{tikzpicture}
{\scriptsize\texttt{thisstringisalsoverylongandisbiggerthanothers}} \\
\begin{tikzpicture} [nodes={text width=0.4cm, text height=0.15cm, align=center, text depth=0.1cm},
         row sep=-\pgflinewidth, column sep=-\pgflinewidth]
         {\scriptsize
   \matrix (hash)[matrix of nodes, nodes={draw, anchor=center}]
         {
       20 & \grayz & \grayz & \grayz & \grayz & \grayz & \grayz & \grayz & \grayz & \grayz & \grayz & $\cdots$ & \grayz \\
    };
    }
\end{tikzpicture}
      \caption{Semi-perfect hashing example}\label{fig:hashing_example}
      \Description[Semi-perfect hashing]{The hash values of several strings according to our hash function}
\end{figure}
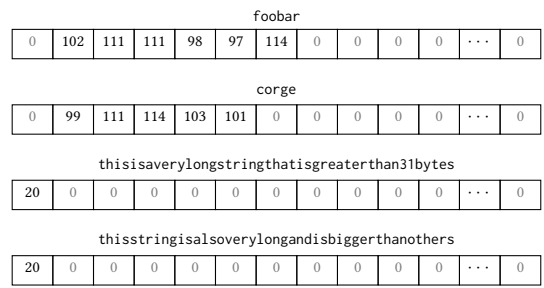

We provide an example of the use of our hash function in Figure~\ref{fig:hashing_example}.
Note that our implementation packs hash values into large integers, but we represent values as byte arrays here for illustrative purposes.
Since the first two strings are less than 32 bytes, they can be compared exactly by only comparing their hash values, avoiding string comparison entirely.
The final two strings are longer than 31 bytes, so we only make use of the first byte, which is calculated based on the string length and the first and last characters.
For these two strings, the hash value is the same since they have the same length and first and last characters.
In this case, as with any hash function, we must compare the entire string in order to check for equality.
However, as discussed previously, we expect this to be rare since most strings we encounter are short.
We provide a detailed evaluation of the benefits of our hash function in Section~\ref{subsubsec:ablation}.

\subsection{Unrolling}\label{subsec:unrolling}

While loop instructions can be useful for flexibility, as with traditional compiler optimization, we find that loop unrolling is sometimes a useful optimization.
There are two cases where we apply unrolling in Blaze: property validation in objects and validation using references.
We describe these cases below along with the heuristics we use to decide when they are employed.
The heuristics were chosen based on observed performance on example test cases.
We leave further tuning of these heuristics to future work.

When validating properties, there are two different approaches that we can take.
The first was previously described in Section~\ref{subsec:loops}.
In this case, we loop over all the properties of an object and look up the appropriate instruction to use for validation based on the property.
However, when most properties are required, it can be more efficient to generate instructions that check each property individually instead.
Specifically, we avoid generating a loop if there are 5 or fewer properties or if at least one quarter of the properties are required.
We also have one additional heuristic that always unrolls loops inside of instructions generated from the \texttt{oneOf} or \texttt{anyOf} applicators.
This increases the likelihood that these operators will be able to quickly short-circuit.

When a part of a schema (typically a definition) is referenced elsewhere, our default approach is to generate instructions for the subschema being referenced and then jump to these instructions as discussed in Section~\ref{subsec:static_ref}.
However, as with loops, these jumps can be detrimental to CPU cache performance.
Therefore, we can instead replace the jump instruction with the necessary instructions to validate according to the referenced subschema.
We perform this replacement if there are no more than 5 references to a particular subschema.
We also avoid this optimization if there are recursive references since they cannot be implemented using unrolling.

\subsection{Regular Expressions}\label{subsec:regexes}

Some features of JSON Schema rely on the evaluation of regular expressions.
This includes, for example, the \texttt{pattern} keyword validating that strings match a particular pattern and \texttt{patternProperties} that serves a similar function for object keys.
We decided to make use of \texttt{Boost.Regex} as our regex engine after observing significantly better performance than \texttt{std::regex} for our use case.
Furthermore, we enable precompilation that trades off time at regex construction for faster matching.
We also identified multiple cases where regular expressions can be further optimized.
For example, many schemas define regular expressions such as \texttt{.*}, that effectively allow all strings.
In addition, expressions such as \texttt{.+} are used to identify non-empty strings.
In both of these cases, we can avoid using a regex engine entirely.
In the first case, where any value is accepted, we can completely remove the regular expression check.
In the second case where the string must be non-empty, we can simply check the length of the string.
Note that the JSON Schema specification does not fully specify the behavior of regular expression matching, so we have chosen to allow \texttt{.} to match any characters.
We expect this will not affect many schemas in practice since line breaks within JSON documents are relatively uncommon.

In a slightly more complicated scenario, regexes such as \texttt{\textasciicircum{}x-} are common in schemas to indicate strings that start with a particular pattern.
(This specific regular expression appeared several hundred times in our corpus of GitHub schemas.)
In this case, we can also avoid using regular expressions by simply checking for a string prefix.
Finally, we also identified a pattern of regular expressions such as \texttt{\textasciicircum.\{3,5\}\$} that effectively indicate a string must be between length 3 and 5.
We can again avoid the use of regular expression matching simply by checking the length of the string.
We implemented special cases for each of these scenarios.
In the future, we plan to explore the conversion of regular expressions to finite automata at compile time to further reduce the matching overhead.

\subsection{Instruction Reordering}\label{subsec:reordering}

As discussed in Section~\ref{sec:compilation}, when compiling to our instruction set from JSON Schema, we have some flexibility on the order instructions are executed.
For complex objects with many properties, we observed that it can be effective to evaluate properties with smaller subschemas first.
For example, 
This has the benefit of potentially allowing validation to fail more quickly while executing fewer instructions.
This optimization is also particularly effective in the presence of applicators such as \texttt{oneOf}.
Failing to validate as quickly as possible in this case means we can more quickly find the correct subschema to validate against.
Note that we currently perform reordering based solely on the size of subschemas.
It is possible that there may be more effective orderings based on the specific data being validated.
For example, we may find that it would be more efficient to place instructions for properties that commonly fail to validate first even if the corresponding subschema is larger.
Currently we take a data-agnostic approach and we leave such further optimizations as future work.

\subsection{Reducing Memory Allocation}

Since our validation happens on the order of nanoseconds in some cases, .
When allocating dynamic data structures such as vectors or hash maps, we prefer to preallocate a small number of entries.
Since most data structures used in our implementation remain quite small, this means we can often avoid further allocations by using this small existing pool.
Furthermore, we optimize for the case of repeated evaluations of the same schema by preallocating a data structure that can be reused for multiple validations without reallocation.
This includes data structures such as pointers to the current location in both the schema and the document being examined.
We plan to explore further memory optimizations in the future such as alternative allocators.

\section{Instruction Execution}\label{sec:execution}

In Blaze, unlike many validator, we do not interpret the schema, but instead precompile it into a set of instructions that can be efficiently executed.
This section describes the Blaze executor.

\subsection{Executor Implementation}

The executor takes a compiled schema as input and executes the instructions against a JSON document to produce a Boolean indicating whether the document is valid.
We first start by describing the structure used to represent each instruction in more detail.
Each instruction first contains the type of instruction, which is one of the values specified in Section~\ref{sec:language} as well as the location in the instance the instruction applies to.
Depending on the instruction, there may also be an associated value such as the expected length of a string or a list of subinstructions used in cases such as validating items in an array according to a set of conditions.

The Blaze executor is driven by a loop over the instructions to be executed.
Each instruction is executed by first looking up the value to be validated from the instance.
Locations in JSON instances are expressed using JSON Pointer notation~\cite{rfc9535} which is used to identify a specific location within a document to be validated.
Note that we make heavy use of the hash function discussed in Section~\ref{subsec:perfect_hashing} to quickly match properties in an object when traversing pointers.
Several instructions may have preconditions to be validated before they are executed. 
For example, the \texttt{Greater} instruction discussed in Section~\ref{subsubsec:assertions} only applies to integers.
In this case, if the corresponding value in the instance being validated is not an integer, the remainder of the instruction will be skipped.
For instructions that contain children, recursion into subinstructions is accomplished by recursively calling the evaluation function in the executor in a loop with each subinstruction.
If any subinstruction fails to validate, the loop over subinstructions is terminated early and failure to validate is returned.
All instance locations specified within instructions are relative to their parent instruction.
We provide a full example of compilation and execution of a schema in the following subsection.

\subsection{Execution Example}

\begin{figure*}[h]
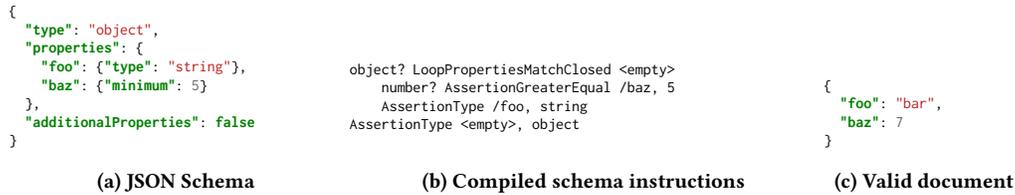

  \centering
  \begin{subfigure}[b]{0.25\textwidth}
      \centering
      \begin{minted}[fontsize=\scriptsize]{json}
{
  "type": "object",
  "properties": {
    "foo": {"type": "string"},
    "baz": {"minimum": 5}
  },
  "additionalProperties": false
}     
      \end{minted} 
      \caption{JSON Schema}\label{fig:example_schema}
      \Description[JSON Schema]{A JSON schema defining a string property foo and a property baz.}
  \end{subfigure}
  \begin{subfigure}[b]{0.35\textwidth}
      \centering
      {\scriptsize
      \begin{verbatim}
object? LoopPropertiesMatchClosed <empty>
    number? AssertionGreaterEqual /baz, 5
    AssertionType /foo, string
AssertionType <empty>, object
      \end{verbatim}
      }
      \caption{Compiled schema instructions}\label{fig:example_instructions}
      \Description[Compiled schema instructions]{The list of instructions compiled from the schema in the previous figure}
  \end{subfigure}
  \begin{subfigure}[b]{0.15\textwidth}
      \centering
      \begin{minted}[fontsize=\scriptsize]{json}
{
  "foo": "bar",
  "baz": 7
}
      \end{minted}
      \caption{Valid document}\label{fig:example_doc}
      \Description[A valid JSON document]{A document containing foo and bar that is valid according to the previous schema}
  \end{subfigure}
  \caption{Example of JSON Schema compilation with Blaze}\label{fig:execution_example}
\end{figure*}

This section provides an example of schema compilation and execution in Blaze using the schema in Figure~\ref{fig:example_schema}.
The schema defines two optional properties \texttt{"foo"} and \texttt{"baz"}.
Each instruction in Figure~\ref{fig:example_instructions} contains both a JSON Pointer and (optionally) an associated value.
Note that here, \texttt{<empty>} refers to the empty JSON pointer, indicating that the root of the document will be validated.
Instructions may have an optional precondition, that is indicated before the instruction with a question mark.
In order to validate the two properties, Blaze first generates the \texttt{Loop\-Properties\-Match\-Closed} instruction.
The \texttt{Closed} variant is selected since \texttt{additionalProperties} is set to \texttt{false} and only the properties specified will be permitted.
This means that while the instruction loops over keys within the JSON object, any key without an associated validation instruction will cause validation to fail.
Also note that this instruction is only executed if the value is of type \texttt{object} since this is a precondition for this instruction.
That is, any values that are not objects, will cause this instruction to be skipped.
Importantly, validation does \emph{not} fail if the precondition for an instruction is not met.

Within the loop are two instructions for validating each of the two properties.
The \texttt{AssertionGreaterEqual} instruction is generated from the \texttt{minimum}.
Since it has the precondition \texttt{number?}, it will only apply to numeric values.
This precondition applies to the instance path \texttt{/baz}.
This means that any non-numeric values will be accepted by this instruction which matches the semantics of the \texttt{minimum} keyword.
If the precondition is valid (the value at \texttt{/baz} is a number), then the instruction checks if this value is at least 5.
The second instruction in the loop is \texttt{AssertionType}.
Since this instruction has no precondition, it applies to all values.
Specifically, this instruction validates that the value at the instance path \texttt{/foo} has type \texttt{string}.
Note that the order of the two instructions within the loop are unimportant since Blaze will look up the appropriate instruction based on the associated instance path.

Note that since the \texttt{Loop\-Properties\-Match\-Closed} instruction included a precondition that the value is of type \texttt{object}, that before the final instruction, any non-object values would pass validation.
However, since the schema specifies that the type of the value must be \texttt{object}, Blaze emits one final instruction.
\texttt{AssertionType} here behaves the same as above and validates that the entire value is an object.
Assuming that this assertion evaluates to \texttt{true}, the entire document is considered valid.

Finally, we briefly walk through the execution of these instructions against the schema in Figure~\ref{fig:example_doc}.
The first instruction, \texttt{Loop\-Properties\-Match\-Closed} begins execution since our document meets the precondition of being an object.
This instruction then loops over the all the key-value pairs in the object.
The first key, \texttt{"foo"} corresponds to the \texttt{AssertionType} instruction.
Since the value, \texttt{"baz"} is a string, it passes validation.
The second key, \texttt{"baz"} corresponds to the \texttt{AssertionGreaterEqual} instruction.
The value is a number, so it passes the precondition, and it is greater than 5, so this instruction also evaluates to \texttt{true}.
Note that at this point, if the document had any other keys, the \texttt{Loop\-Properties\-Match\-Closed} instruction would fail to validate since no additional properties are allowed.
However, since our document only has two properties, this instruction evaluates to \texttt{true} and we move on to the final instruction.
Since our document is an object, this instruction also evaluates to \texttt{true} and the entire document is considered valid.

\section{Evaluation}\label{sec:evaluation}

We have two main focuses with our experimental evaluation: validating the correctness of our implementation and comparing the validation performance with existing JSON Schema validators.

\subsection{Correctness}\label{subsec:eval_correct}

In order to experimentally verify the correctness of our implementation, we make use of the official JSON Schema Test Suite~\cite{TestSuite}.
This test suite is maintained by the authors of the JSON Schema specification and is designed to exercise corner cases in JSON Schema validation and contains several hundred tests for each version of the specification for a total of over 6,000 tests.
These tests verify that each keyword is implemented correctly and also test interactions between relevant keywords.
However, we note that the test suite does not cover all possible cases.
Indeed, we later discuss some implementations that pass this test suite but produce errors in our evaluation.
Through our evaluation of Blaze and other implementations, we were also able to discover and report bugs in other JSON Schema validators not captured by the official test suite.
More than 20 different JSON Schema validators publish ongoing reports via Bowtie~\cite{Bowtie}, a system that automatically runs the test suite against the latest version of supported implementations.
Our implementation is one of only twelve that achieves a perfect score on this test suite for the 2020-12 dialect, confirming the correctness of our compilation process.
In addition to this test suite, we also have over 16,000 lines of manually written test cases across the different dialects supported by Blaze.

\subsection{Performance}

In order to test the performance of our validator, we need a number of schemas as well as documents corresponding to each schema.
We collected our schemas from the JSON Schema Store~\footnote{\url{https://www.schemastore.org/json/}}, a repository of JSON Schemas for various configuration file formats.
All of the schemas make use of dialect 7 of the JSON Schema specification with the exception of \texttt{cql2} and \texttt{openapi} which use the 2020-12 dialect.
For several of these schemas, there is a convention used to name files which are designed to be valid according to the schema.
For example, files that use the \texttt{babelrc} schema are typically named \texttt{.babelrc} or \texttt{babelrc.json}.
We use these file names to search for matching files on open source projects on GitHub.
After finding a set of files, we validate them according to the corresponding schema in order to ensure that each document indeed matches the schema.
A summary of the schemas, their size, and the number and size of documents collected is shown in Table~\ref{tab:datasets}.
Note that to measure the size of each schema, we exclude keywords such as \texttt{description} that have no effect on validation.
All datasets are available in our benchmark repository\footnote{\url{https://github.com/sourcemeta-research/jsonschema-benchmark}}.
We also make available measurements for several other implementations that did not meet our selection criteria.
We note that none of these implementations are faster than Blaze on any of our test datasets.
Experiments are run on a machine equipped with two 8-core 2.10 GHz Intel Xeon Silver 4110.
We note that while multiple cores our available, our implementation of Blaze is currently single-threaded.
We leave the possibility of optimizing for parallel execution as future work.

\begin{table}[h]
    {\scriptsize
    \centering
    \begin{tabular}{l r r r}
        \hline
        Name & \# Docs & Schema Size (KB) & Avg. Doc. Size (B) \\
        \hline
        ansible-meta & 333 & 36.1 & 312 \\
        aws-cdk & 483 & 0.7 & 1145 \\
        babelrc & 794 & 6.5 & 140 \\
        clang-format & 133 & 54.2 & 336 \\
        cmake-presets & 967 & 84.0 & 2721 \\
        code-climate & 2484 & 5.9 & 282 \\
        cql2 & 109 & 17.9 & 125 \\
        cspell & 981 & 125.6 & 817 \\
        cypress & 981 & 16.0 & 401 \\
        deno & 987 & 22.4 & 1018 \\
        dependabot & 967 & 9.4 & 403 \\
        draft-04 & 563 & 4.0 & 12631 \\
        fabric-mod & 911 & 11.1 & 691 \\
        geojson & 500 & 45.0 & 52433 \\
        gitpod-configuration & 986 & 13.1 & 354 \\
        helm-chart-lock & 3888 & 1.5 & 342 \\
        importmap & 964 & 0.6 & 630 \\
        jasmine & 980 & 3.6 & 133 \\
        jsconfig & 981 & 59.5 & 177 \\
        jshintrc & 966 & 11.8 & 429 \\
        krakend & 47 & 377.7 & 2431 \\
        lazygit & 280 & 87.8 & 276 \\
        lerna & 985 & 4.6 & 172 \\
        nest-cli & 1025 & 18.9 & 290 \\
        omnisharp & 987 & 13.5 & 595 \\
        openapi & 107 & 32.5 & 165548 \\
        pre-commit-hooks & 985 & 9.6 & 549 \\
        pulumi & 3807 & 7.7 & 251 \\
        semantic-release & 794 & 3.3 & 460 \\
        stale & 961 & 3.7 & 466 \\
        stylecop & 983 & 11.5 & 567 \\
        tmuxinator & 382 & 4.4 & 628 \\
        ui5 & 942 & 94.1 & 487 \\
        ui5-manifest & 611 & 383.5 & 2356 \\
        unreal-engine-uproject & 859 & 10.6 & 394 \\
        vercel & 710 & 37.2 & 406 \\
        yamllint & 984 & 25.5 & 351 \\
    \end{tabular}
    }
    \caption{Datasets used for validator evaluation}\label{tab:datasets}
\end{table}

We compare against a wide variety of validators across multiple programming languages.
Validators were selected based on those available in the Bowtie test system that either pass the entire JSON Schema Test Suite or are significantly popular.
Note that we exclude the implementations dev.harrel.json-schema and 
io.openapi\-processor.json-schema-validator since they performed an order of magnitude slower than other validators in the majority of cases.
We have also included ajv and the Python jsonschema package since these are both very commonly used, despite producing incorrect results in some cases.
A summary of all the implementations we compare with is in Table~\ref{tab:implementations}.
We run each implementation five times on each dataset and measure the time to compile the schema as well as the time for validating all instances.

\begin{table}[h]
    {%
    \centering
    \begin{tabular}{|c|c|c|c|c|c|}
        \hline
        Implementation & Lang. & Version & Correct & AOT & Stars \\ \hline\hline
        \textbf{Blaze (Ours)} & C++ & 1.0.0 & \cmark & \cmark & <100 \\ \hline
        ajv & JS & 6.12.6 & \xmark & \cmark & >10K \\ \hline
        Boon & Rust & 0.6 & \cmark & \cmark & <100 \\ \hline
        Corvus & C\# & 4.0.12 & \cmark & \cmark & $\sim$100 \\ \hline
        jsonschema & Go & 6.0.1 & \cmark & \cmark & $\sim$1K \\ \hline
        jsonschema & Python & 4.23.0 & \xmark & \xmark & $\sim$5K \\ \hline
        JsonSchema.NET & C\# & 7.2.3 & \xmark & \cmark & $\sim$1K \\ \hline
        JSV & Elixir & 0.2.0 & \cmark & \xmark & <10 \\ \hline
        KMP & Kotlin & 0.3.0 & \cmark & \xmark & <100 \\ \hline
        NetworkNT & Java & 1.5.3 & \cmark & \xmark & $\sim$900 \\ \hline
        json\_schemer & Ruby & 2.3.0 & \cmark & \xmark & $\sim$400 \\ \hline
    \end{tabular}
    }
    \caption{Implementation details for each validator}\label{tab:implementations}
\end{table}

\subsubsection{Compilation}

As previously noted, Blaze trades off an upfront period of compilation for faster validation at runtime.
There are existing validators with a precompilation step, which we indicate in Table~\ref{tab:implementations}.
However, many of these validators have very basic precompilation compared to Blaze.
In several cases, precompilation consists of only parsing and validating the schema itself.

\begin{figure}
    \includeinkscape[width=0.4\textwidth]{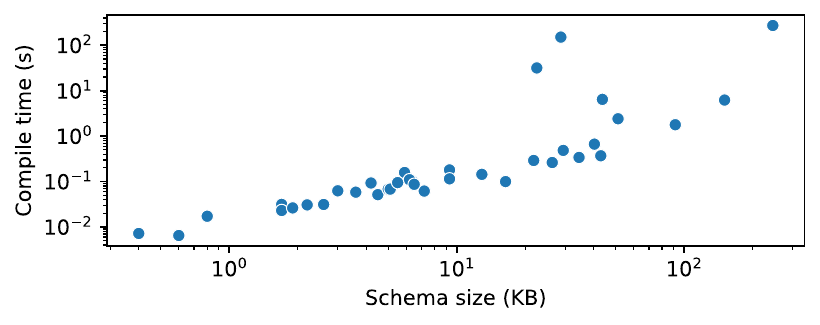}
    \caption{Blaze compilation time relative to schema size}\label{fig:compilation_time}
    \Description[Compilation time]{A plot of compilation time vs the size of each schema. There is a general trend that larger schemas take longer to compile.}
\end{figure}

We performed compilation five times on each of the schemas in our dataset and report the average in Figure~\ref{fig:compilation_time}.
As expected, we can see that the compilation time tends to increase relative to the size of the schema.
We note that since compilation is done for the purpose of speeding up evaluation, validators which are slower to compile may achieve a return on this investment after validating a sufficient number of documents.
We have also not currently made any effort to optimize the compilation time of Blaze.

\subsubsection{Validation}

When measuring the validation runtime, we first measure the runtime immediately after compiling.
We then perform a minimum of 100 more iterations of validation in order to warm up the implementation.
Depending on the implementation, this will have the effect of warming CPU caches, triggering JIT compilation, and other side effects that generally result in warm runs being faster.
Warm runs reflect real-world situations such as an API gateway that validates a large number of incoming payloads according to a fixed schema.
Both warm and cold runtimes for all implementations and datasets are reported in Table~\ref{tab:benchmark_runtime}.

\begin{table*}[h]
    \centering
    {\scriptsize 
    \begin{tabular}{|c|c|c|c|c|c|c|c|c|c|c|c|}
        \hline
        \textbf{Dataset} & 
        \textbf{ajv} & \textbf{Blaze} & \textbf{Boon} & \textbf{Corvus} & \textbf{json\_schemer} & \textbf{jsonschema (Go)} & \textbf{jsonschema (Py)} & \textbf{JsonSchema.Net} & \textbf{JSV} & \textbf{KMP} & \textbf{NetworkNT} \\\hline
        \multirow{2}{*}{ansible-meta} & \multicolumn{1}{c|}{21.4} & \multicolumn{1}{c|}{\textbf{0.5}} & \multicolumn{1}{c|}{26.1} & \multicolumn{1}{c|}{439.1} & \multicolumn{1}{c|}{370.8} & \multicolumn{1}{c|}{66.9} & \multicolumn{1}{c|}{846.1} & \multicolumn{1}{c|}{780.3} & \multicolumn{1}{c|}{82.0} & \multicolumn{1}{c|}{175.0} & \multicolumn{1}{c|}{188.8}\\\cline{2-12} & \multicolumn{1}{c|}{1.4} & \multicolumn{1}{c|}{\textbf{0.5}} & \multicolumn{1}{c|}{22.8} & \multicolumn{1}{c|}{14.0} & \multicolumn{1}{c|}{344.1} & \multicolumn{1}{c|}{68.4} & \multicolumn{1}{c|}{921.9} & \multicolumn{1}{c|}{190.3} & \multicolumn{1}{c|}{63.8} & \multicolumn{1}{c|}{26.8} & \multicolumn{1}{c|}{7.9}\\\hline
        \multirow{2}{*}{aws-cdk} & \multicolumn{1}{c|}{1.8} & \multicolumn{1}{c|}{\textbf{0.1}} & \multicolumn{1}{c|}{0.5} & \multicolumn{1}{c|}{30.2} & \multicolumn{1}{c|}{41.2} & \multicolumn{1}{c|}{4.2} & \multicolumn{1}{c|}{27.1} & \multicolumn{1}{c|}{178.4} & \multicolumn{1}{c|}{23.7} & \multicolumn{1}{c|}{43.8} & \multicolumn{1}{c|}{17.4}\\\cline{2-12} & \multicolumn{1}{c|}{0.2} & \multicolumn{1}{c|}{\textbf{0.1}} & \multicolumn{1}{c|}{0.4} & \multicolumn{1}{c|}{0.5} & \multicolumn{1}{c|}{25.7} & \multicolumn{1}{c|}{3.3} & \multicolumn{1}{c|}{26.0} & \multicolumn{1}{c|}{5.8} & \multicolumn{1}{c|}{6.4} & \multicolumn{1}{c|}{0.8} & \multicolumn{1}{c|}{0.2}\\\hline
        \multirow{2}{*}{babelrc} & \multicolumn{1}{c|}{6.9} & \multicolumn{1}{c|}{\textbf{0.3}} & \multicolumn{1}{c|}{0.8} & \multicolumn{1}{c|}{63.8} & \multicolumn{1}{c|}{73.8} & \multicolumn{1}{c|}{11.2} & \multicolumn{1}{c|}{92.4} & \multicolumn{1}{c|}{257.2} & \multicolumn{1}{c|}{16.1} & \multicolumn{1}{c|}{71.6} & \multicolumn{1}{c|}{28.8}\\\cline{2-12} & \multicolumn{1}{c|}{0.3} & \multicolumn{1}{c|}{\textbf{0.2}} & \multicolumn{1}{c|}{0.7} & \multicolumn{1}{c|}{1.5} & \multicolumn{1}{c|}{57.4} & \multicolumn{1}{c|}{9.2} & \multicolumn{1}{c|}{89.9} & \multicolumn{1}{c|}{21.3} & \multicolumn{1}{c|}{7.3} & \multicolumn{1}{c|}{5.6} & \multicolumn{1}{c|}{1.0}\\\hline
        \multirow{2}{*}{clang-format} & \multicolumn{1}{c|}{15.5} & \multicolumn{1}{c|}{\textbf{0.2}} & \multicolumn{1}{c|}{0.3} & \multicolumn{1}{c|}{338.7} & \multicolumn{1}{c|}{20.2} & \multicolumn{1}{c|}{3.5} & \multicolumn{1}{c|}{27.5} & \multicolumn{1}{c|}{310.7} & \multicolumn{1}{c|}{14.3} & \multicolumn{1}{c|}{42.0} & \multicolumn{1}{c|}{160.2}\\\cline{2-12} & \multicolumn{1}{c|}{1.0} & \multicolumn{1}{c|}{\textbf{0.2}} & \multicolumn{1}{c|}{0.3} & \multicolumn{1}{c|}{9.2} & \multicolumn{1}{c|}{17.0} & \multicolumn{1}{c|}{3.3} & \multicolumn{1}{c|}{19.8} & \multicolumn{1}{c|}{24.9} & \multicolumn{1}{c|}{4.6} & \multicolumn{1}{c|}{0.5} & \multicolumn{1}{c|}{3.9}\\\hline
        \multirow{2}{*}{cmake-presets} & \multicolumn{1}{c|}{279.3} & \multicolumn{1}{c|}{\textbf{12.2}} & \multicolumn{1}{c|}{260.9} & \multicolumn{1}{c|}{673.7} & \multicolumn{1}{c|}{13893.5} & \multicolumn{1}{c|}{744.9} & \multicolumn{1}{c|}{24338.0} & \multicolumn{1}{c|}{11515.3} & \multicolumn{1}{c|}{8185.7} & \multicolumn{1}{c|}{1034.2} & \multicolumn{1}{c|}{1343.1}\\\cline{2-12} & \multicolumn{1}{c|}{118.2} & \multicolumn{1}{c|}{\textbf{8.0}} & \multicolumn{1}{c|}{264.4} & \multicolumn{1}{c|}{97.0} & \multicolumn{1}{c|}{13591.5} & \multicolumn{1}{c|}{751.6} & \multicolumn{1}{c|}{24114.8} & \multicolumn{1}{c|}{9151.3} & \multicolumn{1}{c|}{5969.2} & \multicolumn{1}{c|}{500.4} & \multicolumn{1}{c|}{743.4}\\\hline
        \multirow{2}{*}{code-climate} & \multicolumn{1}{c|}{\textdagger} & \multicolumn{1}{c|}{\textbf{0.6}} & \multicolumn{1}{c|}{1.9} & \multicolumn{1}{c|}{34.6} & \multicolumn{1}{c|}{132.9} & \multicolumn{1}{c|}{8.9} & \multicolumn{1}{c|}{124.5} & \multicolumn{1}{c|}{373.5} & \multicolumn{1}{c|}{21.4} & \multicolumn{1}{c|}{67.5} & \multicolumn{1}{c|}{29.3}\\\cline{2-12} & \multicolumn{1}{c|}{\textdagger} & \multicolumn{1}{c|}{\textbf{0.4}} & \multicolumn{1}{c|}{1.2} & \multicolumn{1}{c|}{2.0} & \multicolumn{1}{c|}{118.7} & \multicolumn{1}{c|}{5.6} & \multicolumn{1}{c|}{104.7} & \multicolumn{1}{c|}{51.6} & \multicolumn{1}{c|}{8.7} & \multicolumn{1}{c|}{2.8} & \multicolumn{1}{c|}{1.2}\\\hline
        \multirow{2}{*}{cql2} & \multicolumn{1}{c|}{87.7} & \multicolumn{1}{c|}{\textbf{0.6}} & \multicolumn{1}{c|}{794.7} & \multicolumn{1}{c|}{315.3} & \multicolumn{1}{c|}{8607.0} & \multicolumn{1}{c|}{375.3} & \multicolumn{1}{c|}{43168.3} & \multicolumn{1}{c|}{\textdagger} & \multicolumn{1}{c|}{\textdagger} & \multicolumn{1}{c|}{683.6} & \multicolumn{1}{c|}{3911.7}\\\cline{2-12} & \multicolumn{1}{c|}{15.5} & \multicolumn{1}{c|}{\textbf{0.4}} & \multicolumn{1}{c|}{762.4} & \multicolumn{1}{c|}{8.2} & \multicolumn{1}{c|}{7255.4} & \multicolumn{1}{c|}{365.7} & \multicolumn{1}{c|}{40420.3} & \multicolumn{1}{c|}{\textdagger} & \multicolumn{1}{c|}{\textdagger} & \multicolumn{1}{c|}{349.3} & \multicolumn{1}{c|}{697.0}\\\hline
        \multirow{2}{*}{cspell} & \multicolumn{1}{c|}{17.9} & \multicolumn{1}{c|}{\textbf{1.9}} & \multicolumn{1}{c|}{\textdagger} & \multicolumn{1}{c|}{258.1} & \multicolumn{1}{c|}{\textdagger} & \multicolumn{1}{c|}{\textdagger} & \multicolumn{1}{c|}{1049.7} & \multicolumn{1}{c|}{1089.2} & \multicolumn{1}{c|}{109.7} & \multicolumn{1}{c|}{\textdagger} & \multicolumn{1}{c|}{314.4}\\\cline{2-12} & \multicolumn{1}{c|}{2.5} & \multicolumn{1}{c|}{\textbf{1.6}} & \multicolumn{1}{c|}{\textdagger} & \multicolumn{1}{c|}{7.1} & \multicolumn{1}{c|}{\textdagger} & \multicolumn{1}{c|}{\textdagger} & \multicolumn{1}{c|}{1075.4} & \multicolumn{1}{c|}{225.0} & \multicolumn{1}{c|}{87.6} & \multicolumn{1}{c|}{\textdagger} & \multicolumn{1}{c|}{13.2}\\\hline
        \multirow{2}{*}{cypress} & \multicolumn{1}{c|}{12.8} & \multicolumn{1}{c|}{\textbf{0.3}} & \multicolumn{1}{c|}{1.1} & \multicolumn{1}{c|}{127.6} & \multicolumn{1}{c|}{74.9} & \multicolumn{1}{c|}{7.5} & \multicolumn{1}{c|}{81.7} & \multicolumn{1}{c|}{395.9} & \multicolumn{1}{c|}{18.2} & \multicolumn{1}{c|}{78.1} & \multicolumn{1}{c|}{28.5}\\\cline{2-12} & \multicolumn{1}{c|}{1.6} & \multicolumn{1}{c|}{\textbf{0.3}} & \multicolumn{1}{c|}{0.9} & \multicolumn{1}{c|}{3.2} & \multicolumn{1}{c|}{67.1} & \multicolumn{1}{c|}{8.9} & \multicolumn{1}{c|}{78.7} & \multicolumn{1}{c|}{36.8} & \multicolumn{1}{c|}{8.6} & \multicolumn{1}{c|}{6.8} & \multicolumn{1}{c|}{1.1}\\\hline
        \multirow{2}{*}{deno} & \multicolumn{1}{c|}{12.9} & \multicolumn{1}{c|}{\textbf{1.1}} & \multicolumn{1}{c|}{1.8} & \multicolumn{1}{c|}{174.7} & \multicolumn{1}{c|}{151.7} & \multicolumn{1}{c|}{14.7} & \multicolumn{1}{c|}{109.2} & \multicolumn{1}{c|}{412.2} & \multicolumn{1}{c|}{49.6} & \multicolumn{1}{c|}{83.4} & \multicolumn{1}{c|}{350.6}\\\cline{2-12} & \multicolumn{1}{c|}{1.4} & \multicolumn{1}{c|}{\textbf{1.0}} & \multicolumn{1}{c|}{1.4} & \multicolumn{1}{c|}{8.0} & \multicolumn{1}{c|}{127.9} & \multicolumn{1}{c|}{14.1} & \multicolumn{1}{c|}{114.4} & \multicolumn{1}{c|}{58.5} & \multicolumn{1}{c|}{27.6} & \multicolumn{1}{c|}{6.2} & \multicolumn{1}{c|}{17.3}\\\hline
        \multirow{2}{*}{dependabot} & \multicolumn{1}{c|}{5.9} & \multicolumn{1}{c|}{\textbf{0.8}} & \multicolumn{1}{c|}{3.0} & \multicolumn{1}{c|}{109.8} & \multicolumn{1}{c|}{177.4} & \multicolumn{1}{c|}{29.1} & \multicolumn{1}{c|}{209.5} & \multicolumn{1}{c|}{323.9} & \multicolumn{1}{c|}{28.5} & \multicolumn{1}{c|}{109.8} & \multicolumn{1}{c|}{40.9}\\\cline{2-12} & \multicolumn{1}{c|}{\textbf{0.8}} & \multicolumn{1}{c|}{1.0} & \multicolumn{1}{c|}{2.7} & \multicolumn{1}{c|}{7.5} & \multicolumn{1}{c|}{192.2} & \multicolumn{1}{c|}{24.8} & \multicolumn{1}{c|}{193.0} & \multicolumn{1}{c|}{43.4} & \multicolumn{1}{c|}{17.1} & \multicolumn{1}{c|}{6.2} & \multicolumn{1}{c|}{2.6}\\\hline
        \multirow{2}{*}{draft-04} & \multicolumn{1}{c|}{62.0} & \multicolumn{1}{c|}{\textbf{10.7}} & \multicolumn{1}{c|}{33.1} & \multicolumn{1}{c|}{429.8} & \multicolumn{1}{c|}{2167.4} & \multicolumn{1}{c|}{162.3} & \multicolumn{1}{c|}{4201.2} & \multicolumn{1}{c|}{2152.2} & \multicolumn{1}{c|}{\textdagger} & \multicolumn{1}{c|}{288.2} & \multicolumn{1}{c|}{256.4}\\\cline{2-12} & \multicolumn{1}{c|}{23.7} & \multicolumn{1}{c|}{\textbf{10.7}} & \multicolumn{1}{c|}{31.5} & \multicolumn{1}{c|}{153.8} & \multicolumn{1}{c|}{2083.0} & \multicolumn{1}{c|}{140.6} & \multicolumn{1}{c|}{4022.6} & \multicolumn{1}{c|}{590.4} & \multicolumn{1}{c|}{\textdagger} & \multicolumn{1}{c|}{107.5} & \multicolumn{1}{c|}{35.6}\\\hline
        \multirow{2}{*}{fabric-mod} & \multicolumn{1}{c|}{16.6} & \multicolumn{1}{c|}{\textbf{2.0}} & \multicolumn{1}{c|}{5.9} & \multicolumn{1}{c|}{144.1} & \multicolumn{1}{c|}{305.9} & \multicolumn{1}{c|}{57.9} & \multicolumn{1}{c|}{740.0} & \multicolumn{1}{c|}{593.5} & \multicolumn{1}{c|}{171.1} & \multicolumn{1}{c|}{143.4} & \multicolumn{1}{c|}{296.1}\\\cline{2-12} & \multicolumn{1}{c|}{2.4} & \multicolumn{1}{c|}{\textbf{1.9}} & \multicolumn{1}{c|}{4.9} & \multicolumn{1}{c|}{7.7} & \multicolumn{1}{c|}{296.2} & \multicolumn{1}{c|}{54.9} & \multicolumn{1}{c|}{707.1} & \multicolumn{1}{c|}{108.6} & \multicolumn{1}{c|}{103.3} & \multicolumn{1}{c|}{14.7} & \multicolumn{1}{c|}{12.0}\\\hline
        \multirow{2}{*}{geojson} & \multicolumn{1}{c|}{217.9} & \multicolumn{1}{c|}{\textbf{44.3}} & \multicolumn{1}{c|}{1492.5} & \multicolumn{1}{c|}{1127.5} & \multicolumn{1}{c|}{31812.3} & \multicolumn{1}{c|}{2122.7} & \multicolumn{1}{c|}{107120.1} & \multicolumn{1}{c|}{37934.5} & \multicolumn{1}{c|}{28436.3} & \multicolumn{1}{c|}{10501.7} & \multicolumn{1}{c|}{\textdagger}\\\cline{2-12} & \multicolumn{1}{c|}{53.2} & \multicolumn{1}{c|}{\textbf{27.2}} & \multicolumn{1}{c|}{1500.5} & \multicolumn{1}{c|}{196.1} & \multicolumn{1}{c|}{33601.6} & \multicolumn{1}{c|}{2065.6} & \multicolumn{1}{c|}{94707.2} & \multicolumn{1}{c|}{33199.4} & \multicolumn{1}{c|}{19140.4} & \multicolumn{1}{c|}{9381.7} & \multicolumn{1}{c|}{\textdagger}\\\hline
        \multirow{2}{*}{gitpod} & \multicolumn{1}{c|}{10.1} & \multicolumn{1}{c|}{\textbf{0.5}} & \multicolumn{1}{c|}{1.7} & \multicolumn{1}{c|}{102.7} & \multicolumn{1}{c|}{152.0} & \multicolumn{1}{c|}{16.1} & \multicolumn{1}{c|}{140.1} & \multicolumn{1}{c|}{346.9} & \multicolumn{1}{c|}{22.8} & \multicolumn{1}{c|}{90.2} & \multicolumn{1}{c|}{132.4}\\\cline{2-12} & \multicolumn{1}{c|}{0.7} & \multicolumn{1}{c|}{\textbf{0.4}} & \multicolumn{1}{c|}{1.4} & \multicolumn{1}{c|}{3.9} & \multicolumn{1}{c|}{138.7} & \multicolumn{1}{c|}{16.6} & \multicolumn{1}{c|}{134.2} & \multicolumn{1}{c|}{40.8} & \multicolumn{1}{c|}{11.7} & \multicolumn{1}{c|}{8.6} & \multicolumn{1}{c|}{1.7}\\\hline
        \multirow{2}{*}{helm-chart-lock} & \multicolumn{1}{c|}{8.3} & \multicolumn{1}{c|}{\textbf{0.7}} & \multicolumn{1}{c|}{6.3} & \multicolumn{1}{c|}{40.1} & \multicolumn{1}{c|}{428.5} & \multicolumn{1}{c|}{42.5} & \multicolumn{1}{c|}{464.6} & \multicolumn{1}{c|}{460.1} & \multicolumn{1}{c|}{135.6} & \multicolumn{1}{c|}{132.2} & \multicolumn{1}{c|}{45.0}\\\cline{2-12} & \multicolumn{1}{c|}{0.8} & \multicolumn{1}{c|}{\textbf{0.6}} & \multicolumn{1}{c|}{5.4} & \multicolumn{1}{c|}{8.1} & \multicolumn{1}{c|}{400.4} & \multicolumn{1}{c|}{34.8} & \multicolumn{1}{c|}{500.0} & \multicolumn{1}{c|}{77.8} & \multicolumn{1}{c|}{95.4} & \multicolumn{1}{c|}{17.4} & \multicolumn{1}{c|}{5.3}\\\hline
        \multirow{2}{*}{importmap} & \multicolumn{1}{c|}{2.9} & \multicolumn{1}{c|}{\textbf{0.1}} & \multicolumn{1}{c|}{1.4} & \multicolumn{1}{c|}{28.3} & \multicolumn{1}{c|}{94.0} & \multicolumn{1}{c|}{12.1} & \multicolumn{1}{c|}{84.8} & \multicolumn{1}{c|}{218.2} & \multicolumn{1}{c|}{33.1} & \multicolumn{1}{c|}{59.8} & \multicolumn{1}{c|}{24.3}\\\cline{2-12} & \multicolumn{1}{c|}{0.3} & \multicolumn{1}{c|}{\textbf{0.1}} & \multicolumn{1}{c|}{1.2} & \multicolumn{1}{c|}{0.8} & \multicolumn{1}{c|}{63.4} & \multicolumn{1}{c|}{12.4} & \multicolumn{1}{c|}{86.8} & \multicolumn{1}{c|}{14.3} & \multicolumn{1}{c|}{18.8} & \multicolumn{1}{c|}{1.6} & \multicolumn{1}{c|}{1.1}\\\hline
        \multirow{2}{*}{jasmine} & \multicolumn{1}{c|}{3.2} & \multicolumn{1}{c|}{\textbf{0.3}} & \multicolumn{1}{c|}{1.2} & \multicolumn{1}{c|}{48.3} & \multicolumn{1}{c|}{90.5} & \multicolumn{1}{c|}{14.4} & \multicolumn{1}{c|}{138.3} & \multicolumn{1}{c|}{279.1} & \multicolumn{1}{c|}{16.8} & \multicolumn{1}{c|}{97.1} & \multicolumn{1}{c|}{25.2}\\\cline{2-12} & \multicolumn{1}{c|}{0.3} & \multicolumn{1}{c|}{\textbf{0.2}} & \multicolumn{1}{c|}{1.1} & \multicolumn{1}{c|}{2.0} & \multicolumn{1}{c|}{75.6} & \multicolumn{1}{c|}{13.6} & \multicolumn{1}{c|}{132.9} & \multicolumn{1}{c|}{23.1} & \multicolumn{1}{c|}{8.8} & \multicolumn{1}{c|}{6.6} & \multicolumn{1}{c|}{1.1}\\\hline
        \multirow{2}{*}{jsconfig} & \multicolumn{1}{c|}{17.8} & \multicolumn{1}{c|}{\textbf{1.2}} & \multicolumn{1}{c|}{3.7} & \multicolumn{1}{c|}{285.9} & \multicolumn{1}{c|}{188.1} & \multicolumn{1}{c|}{34.5} & \multicolumn{1}{c|}{356.9} & \multicolumn{1}{c|}{5019.7} & \multicolumn{1}{c|}{43.7} & \multicolumn{1}{c|}{160.2} & \multicolumn{1}{c|}{325.0}\\\cline{2-12} & \multicolumn{1}{c|}{3.1} & \multicolumn{1}{c|}{\textbf{1.1}} & \multicolumn{1}{c|}{2.9} & \multicolumn{1}{c|}{9.1} & \multicolumn{1}{c|}{183.3} & \multicolumn{1}{c|}{29.3} & \multicolumn{1}{c|}{349.2} & \multicolumn{1}{c|}{4340.1} & \multicolumn{1}{c|}{33.5} & \multicolumn{1}{c|}{19.2} & \multicolumn{1}{c|}{8.5}\\\hline
        \multirow{2}{*}{jshintrc} & \multicolumn{1}{c|}{9.1} & \multicolumn{1}{c|}{\textbf{1.6}} & \multicolumn{1}{c|}{2.5} & \multicolumn{1}{c|}{173.9} & \multicolumn{1}{c|}{145.1} & \multicolumn{1}{c|}{24.4} & \multicolumn{1}{c|}{142.7} & \multicolumn{1}{c|}{320.0} & \multicolumn{1}{c|}{37.3} & \multicolumn{1}{c|}{78.3} & \multicolumn{1}{c|}{35.5}\\\cline{2-12} & \multicolumn{1}{c|}{2.6} & \multicolumn{1}{c|}{\textbf{1.6}} & \multicolumn{1}{c|}{2.5} & \multicolumn{1}{c|}{19.7} & \multicolumn{1}{c|}{141.1} & \multicolumn{1}{c|}{20.0} & \multicolumn{1}{c|}{149.8} & \multicolumn{1}{c|}{38.8} & \multicolumn{1}{c|}{26.3} & \multicolumn{1}{c|}{5.5} & \multicolumn{1}{c|}{2.1}\\\hline
        \multirow{2}{*}{krakend} & \multicolumn{1}{c|}{\textdagger} & \multicolumn{1}{c|}{\textbf{1.0}} & \multicolumn{1}{c|}{1.4} & \multicolumn{1}{c|}{646.8} & \multicolumn{1}{c|}{95.6} & \multicolumn{1}{c|}{12.7} & \multicolumn{1}{c|}{123.5} & \multicolumn{1}{c|}{431.0} & \multicolumn{1}{c|}{39.3} & \multicolumn{1}{c|}{\textdagger} & \multicolumn{1}{c|}{\textdagger}\\\cline{2-12} & \multicolumn{1}{c|}{\textdagger} & \multicolumn{1}{c|}{\textbf{0.6}} & \multicolumn{1}{c|}{1.0} & \multicolumn{1}{c|}{6.9} & \multicolumn{1}{c|}{70.4} & \multicolumn{1}{c|}{10.6} & \multicolumn{1}{c|}{121.1} & \multicolumn{1}{c|}{34.4} & \multicolumn{1}{c|}{29.1} & \multicolumn{1}{c|}{\textdagger} & \multicolumn{1}{c|}{\textdagger}\\\hline
        \multirow{2}{*}{lazygit} & \multicolumn{1}{c|}{33.7} & \multicolumn{1}{c|}{\textbf{0.5}} & \multicolumn{1}{c|}{0.9} & \multicolumn{1}{c|}{359.9} & \multicolumn{1}{c|}{78.5} & \multicolumn{1}{c|}{6.2} & \multicolumn{1}{c|}{87.7} & \multicolumn{1}{c|}{396.0} & \multicolumn{1}{c|}{19.9} & \multicolumn{1}{c|}{75.8} & \multicolumn{1}{c|}{203.0}\\\cline{2-12} & \multicolumn{1}{c|}{1.9} & \multicolumn{1}{c|}{\textbf{0.3}} & \multicolumn{1}{c|}{0.7} & \multicolumn{1}{c|}{5.8} & \multicolumn{1}{c|}{53.6} & \multicolumn{1}{c|}{7.4} & \multicolumn{1}{c|}{85.9} & \multicolumn{1}{c|}{38.1} & \multicolumn{1}{c|}{6.9} & \multicolumn{1}{c|}{4.8} & \multicolumn{1}{c|}{2.1}\\\hline
        \multirow{2}{*}{lerna} & \multicolumn{1}{c|}{3.9} & \multicolumn{1}{c|}{\textbf{0.4}} & \multicolumn{1}{c|}{0.9} & \multicolumn{1}{c|}{51.7} & \multicolumn{1}{c|}{67.5} & \multicolumn{1}{c|}{7.8} & \multicolumn{1}{c|}{52.8} & \multicolumn{1}{c|}{247.2} & \multicolumn{1}{c|}{16.4} & \multicolumn{1}{c|}{58.5} & \multicolumn{1}{c|}{21.9}\\\cline{2-12} & \multicolumn{1}{c|}{0.4} & \multicolumn{1}{c|}{\textbf{0.3}} & \multicolumn{1}{c|}{0.8} & \multicolumn{1}{c|}{1.4} & \multicolumn{1}{c|}{53.5} & \multicolumn{1}{c|}{8.8} & \multicolumn{1}{c|}{56.0} & \multicolumn{1}{c|}{17.5} & \multicolumn{1}{c|}{6.5} & \multicolumn{1}{c|}{1.9} & \multicolumn{1}{c|}{0.9}\\\hline
        \multirow{2}{*}{nest-cli} & \multicolumn{1}{c|}{7.2} & \multicolumn{1}{c|}{\textbf{0.5}} & \multicolumn{1}{c|}{2.0} & \multicolumn{1}{c|}{103.9} & \multicolumn{1}{c|}{126.3} & \multicolumn{1}{c|}{14.8} & \multicolumn{1}{c|}{186.0} & \multicolumn{1}{c|}{440.8} & \multicolumn{1}{c|}{22.3} & \multicolumn{1}{c|}{100.0} & \multicolumn{1}{c|}{37.7}\\\cline{2-12} & \multicolumn{1}{c|}{0.7} & \multicolumn{1}{c|}{\textbf{0.4}} & \multicolumn{1}{c|}{1.7} & \multicolumn{1}{c|}{3.3} & \multicolumn{1}{c|}{145.9} & \multicolumn{1}{c|}{15.0} & \multicolumn{1}{c|}{190.0} & \multicolumn{1}{c|}{53.9} & \multicolumn{1}{c|}{13.9} & \multicolumn{1}{c|}{8.2} & \multicolumn{1}{c|}{1.4}\\\hline
        \multirow{2}{*}{omnisharp} & \multicolumn{1}{c|}{10.6} & \multicolumn{1}{c|}{\textbf{1.3}} & \multicolumn{1}{c|}{2.0} & \multicolumn{1}{c|}{188.2} & \multicolumn{1}{c|}{107.5} & \multicolumn{1}{c|}{17.8} & \multicolumn{1}{c|}{105.2} & \multicolumn{1}{c|}{325.9} & \multicolumn{1}{c|}{26.5} & \multicolumn{1}{c|}{72.4} & \multicolumn{1}{c|}{29.1}\\\cline{2-12} & \multicolumn{1}{c|}{1.5} & \multicolumn{1}{c|}{\textbf{1.2}} & \multicolumn{1}{c|}{2.0} & \multicolumn{1}{c|}{7.0} & \multicolumn{1}{c|}{99.8} & \multicolumn{1}{c|}{13.9} & \multicolumn{1}{c|}{109.0} & \multicolumn{1}{c|}{51.3} & \multicolumn{1}{c|}{15.5} & \multicolumn{1}{c|}{6.4} & \multicolumn{1}{c|}{1.3}\\\hline
        \multirow{2}{*}{openapi} & \multicolumn{1}{c|}{\textdagger} & \multicolumn{1}{c|}{\textbf{59.3}} & \multicolumn{1}{c|}{94.6} & \multicolumn{1}{c|}{540.2} & \multicolumn{1}{c|}{3331.7} & \multicolumn{1}{c|}{300.0} & \multicolumn{1}{c|}{48583.0} & \multicolumn{1}{c|}{4438.6} & \multicolumn{1}{c|}{1944.4} & \multicolumn{1}{c|}{\textdagger} & \multicolumn{1}{c|}{1258.4}\\\cline{2-12} & \multicolumn{1}{c|}{\textdagger} & \multicolumn{1}{c|}{\textbf{27.6}} & \multicolumn{1}{c|}{92.3} & \multicolumn{1}{c|}{72.1} & \multicolumn{1}{c|}{3152.3} & \multicolumn{1}{c|}{266.8} & \multicolumn{1}{c|}{45347.2} & \multicolumn{1}{c|}{2991.3} & \multicolumn{1}{c|}{1506.9} & \multicolumn{1}{c|}{\textdagger} & \multicolumn{1}{c|}{462.0}\\\hline
        \multirow{2}{*}{pre-commit} & \multicolumn{1}{c|}{15.2} & \multicolumn{1}{c|}{\textbf{1.0}} & \multicolumn{1}{c|}{4.2} & \multicolumn{1}{c|}{138.1} & \multicolumn{1}{c|}{210.3} & \multicolumn{1}{c|}{33.6} & \multicolumn{1}{c|}{358.8} & \multicolumn{1}{c|}{432.6} & \multicolumn{1}{c|}{114.9} & \multicolumn{1}{c|}{117.8} & \multicolumn{1}{c|}{39.5}\\\cline{2-12} & \multicolumn{1}{c|}{4.9} & \multicolumn{1}{c|}{\textbf{0.9}} & \multicolumn{1}{c|}{4.0} & \multicolumn{1}{c|}{24.9} & \multicolumn{1}{c|}{184.7} & \multicolumn{1}{c|}{34.6} & \multicolumn{1}{c|}{351.2} & \multicolumn{1}{c|}{84.9} & \multicolumn{1}{c|}{69.9} & \multicolumn{1}{c|}{11.2} & \multicolumn{1}{c|}{4.9}\\\hline
        \multirow{2}{*}{pulumi} & \multicolumn{1}{c|}{14.8} & \multicolumn{1}{c|}{\textbf{1.5}} & \multicolumn{1}{c|}{4.7} & \multicolumn{1}{c|}{99.3} & \multicolumn{1}{c|}{319.8} & \multicolumn{1}{c|}{31.4} & \multicolumn{1}{c|}{396.0} & \multicolumn{1}{c|}{850.7} & \multicolumn{1}{c|}{111.3} & \multicolumn{1}{c|}{144.9} & \multicolumn{1}{c|}{70.6}\\\cline{2-12} & \multicolumn{1}{c|}{2.2} & \multicolumn{1}{c|}{\textbf{1.3}} & \multicolumn{1}{c|}{4.7} & \multicolumn{1}{c|}{6.6} & \multicolumn{1}{c|}{323.7} & \multicolumn{1}{c|}{29.9} & \multicolumn{1}{c|}{411.5} & \multicolumn{1}{c|}{152.0} & \multicolumn{1}{c|}{66.1} & \multicolumn{1}{c|}{15.6} & \multicolumn{1}{c|}{5.4}\\\hline
        \multirow{2}{*}{semantic-release} & \multicolumn{1}{c|}{4.0} & \multicolumn{1}{c|}{\textbf{0.3}} & \multicolumn{1}{c|}{1.8} & \multicolumn{1}{c|}{51.7} & \multicolumn{1}{c|}{117.2} & \multicolumn{1}{c|}{22.5} & \multicolumn{1}{c|}{203.6} & \multicolumn{1}{c|}{353.8} & \multicolumn{1}{c|}{18.8} & \multicolumn{1}{c|}{79.5} & \multicolumn{1}{c|}{45.5}\\\cline{2-12} & \multicolumn{1}{c|}{0.4} & \multicolumn{1}{c|}{\textbf{0.2}} & \multicolumn{1}{c|}{1.6} & \multicolumn{1}{c|}{1.2} & \multicolumn{1}{c|}{103.1} & \multicolumn{1}{c|}{17.2} & \multicolumn{1}{c|}{211.1} & \multicolumn{1}{c|}{45.5} & \multicolumn{1}{c|}{9.9} & \multicolumn{1}{c|}{5.6} & \multicolumn{1}{c|}{1.5}\\\hline
        \multirow{2}{*}{stale} & \multicolumn{1}{c|}{4.2} & \multicolumn{1}{c|}{\textbf{0.3}} & \multicolumn{1}{c|}{1.0} & \multicolumn{1}{c|}{53.9} & \multicolumn{1}{c|}{89.3} & \multicolumn{1}{c|}{13.6} & \multicolumn{1}{c|}{96.6} & \multicolumn{1}{c|}{256.3} & \multicolumn{1}{c|}{\textdagger} & \multicolumn{1}{c|}{72.1} & \multicolumn{1}{c|}{33.6}\\\cline{2-12} & \multicolumn{1}{c|}{0.3} & \multicolumn{1}{c|}{\textbf{0.3}} & \multicolumn{1}{c|}{0.9} & \multicolumn{1}{c|}{1.7} & \multicolumn{1}{c|}{81.4} & \multicolumn{1}{c|}{14.2} & \multicolumn{1}{c|}{98.1} & \multicolumn{1}{c|}{17.8} & \multicolumn{1}{c|}{\textdagger} & \multicolumn{1}{c|}{2.2} & \multicolumn{1}{c|}{1.0}\\\hline
        \multirow{2}{*}{stylecop} & \multicolumn{1}{c|}{6.9} & \multicolumn{1}{c|}{\textbf{1.1}} & \multicolumn{1}{c|}{1.9} & \multicolumn{1}{c|}{131.3} & \multicolumn{1}{c|}{152.0} & \multicolumn{1}{c|}{16.3} & \multicolumn{1}{c|}{151.2} & \multicolumn{1}{c|}{325.9} & \multicolumn{1}{c|}{55.9} & \multicolumn{1}{c|}{95.3} & \multicolumn{1}{c|}{274.1}\\\cline{2-12} & \multicolumn{1}{c|}{1.1} & \multicolumn{1}{c|}{\textbf{0.9}} & \multicolumn{1}{c|}{1.5} & \multicolumn{1}{c|}{4.5} & \multicolumn{1}{c|}{126.8} & \multicolumn{1}{c|}{15.5} & \multicolumn{1}{c|}{139.6} & \multicolumn{1}{c|}{43.4} & \multicolumn{1}{c|}{33.9} & \multicolumn{1}{c|}{6.8} & \multicolumn{1}{c|}{9.9}\\\hline
        \multirow{2}{*}{tmuxinator} & \multicolumn{1}{c|}{4.2} & \multicolumn{1}{c|}{\textbf{0.3}} & \multicolumn{1}{c|}{0.9} & \multicolumn{1}{c|}{47.9} & \multicolumn{1}{c|}{63.6} & \multicolumn{1}{c|}{9.4} & \multicolumn{1}{c|}{105.2} & \multicolumn{1}{c|}{248.9} & \multicolumn{1}{c|}{26.7} & \multicolumn{1}{c|}{60.6} & \multicolumn{1}{c|}{33.6}\\\cline{2-12} & \multicolumn{1}{c|}{0.5} & \multicolumn{1}{c|}{\textbf{0.2}} & \multicolumn{1}{c|}{0.8} & \multicolumn{1}{c|}{1.3} & \multicolumn{1}{c|}{45.3} & \multicolumn{1}{c|}{7.2} & \multicolumn{1}{c|}{111.8} & \multicolumn{1}{c|}{16.9} & \multicolumn{1}{c|}{7.8} & \multicolumn{1}{c|}{2.5} & \multicolumn{1}{c|}{0.8}\\\hline
        \multirow{2}{*}{ui5} & \multicolumn{1}{c|}{111.6} & \multicolumn{1}{c|}{\textbf{1.5}} & \multicolumn{1}{c|}{6.3} & \multicolumn{1}{c|}{719.6} & \multicolumn{1}{c|}{323.8} & \multicolumn{1}{c|}{44.4} & \multicolumn{1}{c|}{443.0} & \multicolumn{1}{c|}{3102.4} & \multicolumn{1}{c|}{46.5} & \multicolumn{1}{c|}{144.2} & \multicolumn{1}{c|}{251.8}\\\cline{2-12} & \multicolumn{1}{c|}{4.7} & \multicolumn{1}{c|}{\textbf{1.1}} & \multicolumn{1}{c|}{5.9} & \multicolumn{1}{c|}{18.3} & \multicolumn{1}{c|}{298.0} & \multicolumn{1}{c|}{39.4} & \multicolumn{1}{c|}{453.3} & \multicolumn{1}{c|}{1479.9} & \multicolumn{1}{c|}{29.2} & \multicolumn{1}{c|}{14.8} & \multicolumn{1}{c|}{8.6}\\\hline
        \multirow{2}{*}{ui5-manifest} & \multicolumn{1}{c|}{\textdagger} & \multicolumn{1}{c|}{\textbf{19.1}} & \multicolumn{1}{c|}{\textdagger} & \multicolumn{1}{c|}{831.0} & \multicolumn{1}{c|}{1172.3} & \multicolumn{1}{c|}{\textdagger} & \multicolumn{1}{c|}{2154.1} & \multicolumn{1}{c|}{2856.8} & \multicolumn{1}{c|}{702.5} & \multicolumn{1}{c|}{315.4} & \multicolumn{1}{c|}{\textdagger}\\\cline{2-12} & \multicolumn{1}{c|}{\textdagger} & \multicolumn{1}{c|}{\textbf{7.3}} & \multicolumn{1}{c|}{\textdagger} & \multicolumn{1}{c|}{69.3} & \multicolumn{1}{c|}{1267.5} & \multicolumn{1}{c|}{\textdagger} & \multicolumn{1}{c|}{2191.0} & \multicolumn{1}{c|}{1144.9} & \multicolumn{1}{c|}{481.7} & \multicolumn{1}{c|}{51.2} & \multicolumn{1}{c|}{\textdagger}\\\hline
        \multirow{2}{*}{unreal} & \multicolumn{1}{c|}{21.3} & \multicolumn{1}{c|}{\textbf{1.1}} & \multicolumn{1}{c|}{2.7} & \multicolumn{1}{c|}{164.4} & \multicolumn{1}{c|}{195.4} & \multicolumn{1}{c|}{27.8} & \multicolumn{1}{c|}{415.4} & \multicolumn{1}{c|}{510.8} & \multicolumn{1}{c|}{83.5} & \multicolumn{1}{c|}{157.4} & \multicolumn{1}{c|}{266.3}\\\cline{2-12} & \multicolumn{1}{c|}{6.4} & \multicolumn{1}{c|}{\textbf{1.0}} & \multicolumn{1}{c|}{2.8} & \multicolumn{1}{c|}{23.2} & \multicolumn{1}{c|}{167.3} & \multicolumn{1}{c|}{24.8} & \multicolumn{1}{c|}{387.0} & \multicolumn{1}{c|}{92.9} & \multicolumn{1}{c|}{36.5} & \multicolumn{1}{c|}{8.2} & \multicolumn{1}{c|}{5.2}\\\hline
        \multirow{2}{*}{vercel} & \multicolumn{1}{c|}{28.3} & \multicolumn{1}{c|}{\textbf{0.6}} & \multicolumn{1}{c|}{2.5} & \multicolumn{1}{c|}{249.8} & \multicolumn{1}{c|}{144.8} & \multicolumn{1}{c|}{18.0} & \multicolumn{1}{c|}{144.3} & \multicolumn{1}{c|}{505.7} & \multicolumn{1}{c|}{65.0} & \multicolumn{1}{c|}{117.7} & \multicolumn{1}{c|}{185.7}\\\cline{2-12} & \multicolumn{1}{c|}{1.5} & \multicolumn{1}{c|}{\textbf{0.6}} & \multicolumn{1}{c|}{1.6} & \multicolumn{1}{c|}{6.6} & \multicolumn{1}{c|}{129.0} & \multicolumn{1}{c|}{13.2} & \multicolumn{1}{c|}{138.8} & \multicolumn{1}{c|}{70.2} & \multicolumn{1}{c|}{25.8} & \multicolumn{1}{c|}{6.6} & \multicolumn{1}{c|}{2.7}\\\hline
        \multirow{2}{*}{yamllint} & \multicolumn{1}{c|}{\textdagger} & \multicolumn{1}{c|}{\textbf{0.1}} & \multicolumn{1}{c|}{0.4} & \multicolumn{1}{c|}{13.3} & \multicolumn{1}{c|}{21.1} & \multicolumn{1}{c|}{3.5} & \multicolumn{1}{c|}{24.7} & \multicolumn{1}{c|}{355.2} & \multicolumn{1}{c|}{\textdagger} & \multicolumn{1}{c|}{38.6} & \multicolumn{1}{c|}{12.8}\\\cline{2-12} & \multicolumn{1}{c|}{\textdagger} & \multicolumn{1}{c|}{\textbf{0.0}} & \multicolumn{1}{c|}{0.3} & \multicolumn{1}{c|}{0.2} & \multicolumn{1}{c|}{20.7} & \multicolumn{1}{c|}{2.0} & \multicolumn{1}{c|}{22.6} & \multicolumn{1}{c|}{55.8} & \multicolumn{1}{c|}{\textdagger} & \multicolumn{1}{c|}{0.5} & \multicolumn{1}{c|}{0.2}\\\hline
    \end{tabular}
}
    \begin{minipage}{0.9\textwidth}
        \smallskip
        {\scriptsize
        The top number is the runtime in milliseconds for the first run after compilation. The bottom is the runtime for validating the same documents after several validation runs. \\
        \textdagger The implementation produced an error on this dataset.
        }
        \bigskip
    \end{minipage}
    \caption{Runtime results for implementations across datsets}\label{tab:benchmark_runtime}
\end{table*}

We note that several implementations produce failures on some schemas despite passing all the tests in the JSON Schema Test Suite.
This is due to some tests being considered optional, which are not counted as failures since they exercise uncommon edge cases.
For example, JSON Pointers that are used for references must be escaped if they contain either slashes or tildes.
Several implementations do not perform this escaping properly, which is necessary for the \texttt{krakend} schema.
Another common failure case is related to regular expressions.
The JSON Schema specification indicates that regular expressions should be interpreted according to the ECMA-262 specification.
Since some languages do not have an ECMA-262-compliant regex engine, some JSON Schema validators choose to use an alternative regex engine.
While most regular expressions used in the schemas in our evaluation are supported across a wide variety of regex engines, there are others that fail to be interpreted correctly.
We plan to explore the possibility of missing test cases in the JSON Schema specification as future work.

\begin{figure}
    \includeinkscape[width=0.4\textwidth]{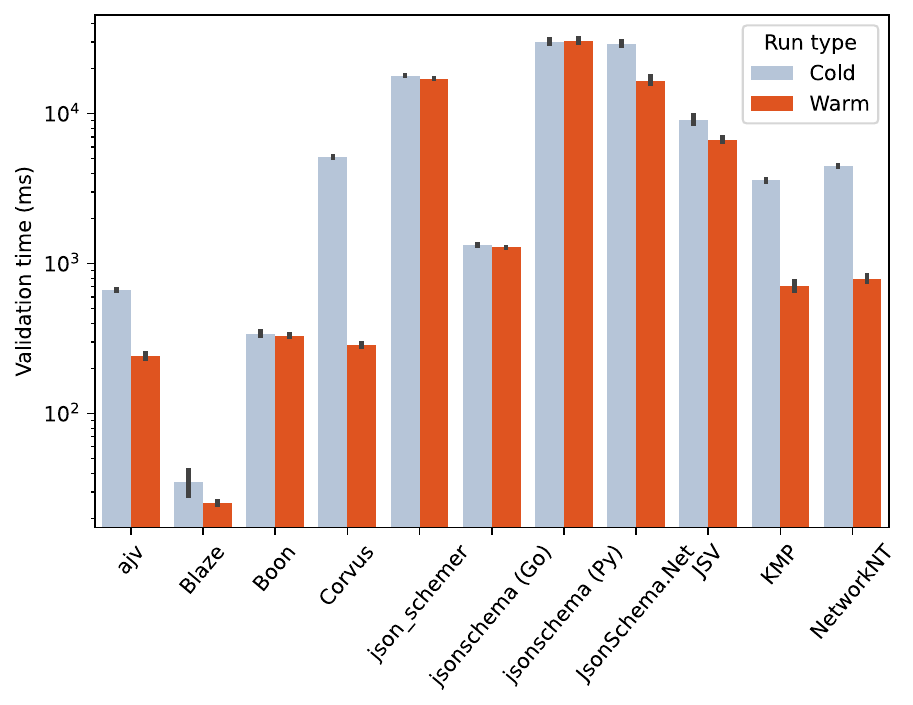}
    \caption{Validation time summary}\label{fig:summary}
    \Description[Validation time]{A plot of validation time across various implementations}
\end{figure}

We show a summary of the performance of Blaze compared to other validators in Figure~\ref{fig:summary} (note the log scale).
For this analysis, we exclude all schemas where any implementation has observed failure and sum the runtimes across the remaining 27 schemas.
We note that Blaze is faster than every other implementation by a minimum of 34\% on every dataset.
Before warmup, Blaze is $\sim$10.9$\times$ faster than the next fastest implementation, Boon.
After warmup, Blaze is approximately $\sim$9.4$\times$ faster than the second fastest implementation, ajv while also being more correct than ajv.

\subsubsection{Optimization}\label{subsubsec:ablation}

In addition to the benchmarks against other validators, we also performed an ablation study to understand the effect of each of our optimizations on the performance of Blaze.
Specifically, we measure the performance with four separate optimizations disabled: our semi-perfect hash function (Section~\ref{subsec:perfect_hashing}), instruction unrolling (Section~\ref{subsec:unrolling}, regular expression optimization (Section~\ref{subsec:regexes}), and instruction reordering (Section~\ref{subsec:reordering}).
We note that while reducing memory allocations was important for the performance of Blaze, this is not an optimization that can easily be disabled since it is integral to our design.
The relative speedup achieved by each optimization is shown for the single schema that is most affected by each optimization as well as the overall runtime for all 38 schemas in our test collection.

Although the average speedup for unrolling is relatively small (only $\sim$3\% overall), we see a benefit of nearly 43\% on one schema, suggesting it can be very effective in certain cases.
We do also note that in the worst case, as with unrolling in compiler optimization, this optimization can reduce performance.
Although Blaze employs heuristics to decide when unrolling is appropriate, we still see a significant performance reduction on several schemas, suggesting a need for improvement in these heuristics in future work.
We also plan to make these optimizations configurable so users can decide which optimizations to use in the case our heuristics are ineffective.

Our regular expression optimizations are similarly effective with an overall improvement of over 10\%.
We do see a minor reduction in runtime in some cases, but in the best case, we again see a runtime reduction of 29\%.
We plan to explore further ways to optimize the use of regular expressions in future work.

We compare the performance of our hash function with the popular and widely-used MurmurHash3\footnote{\url{https://github.com/aappleby/smhasher/wiki/MurmurHash3}}.
Our  function along with instruction reordering are by far the most effective optimizations employed by Blaze, with an average improvement of almost 25\% across all schemas.
In the best case, we see almost a 49\% percent reduction in runtime through the use of our hash function with only a single case where use of our hash function results in a reduction in runtime.
We plan to explore this case further in future work.

\begin{figure*}
    \includeinkscape[width=\textwidth]{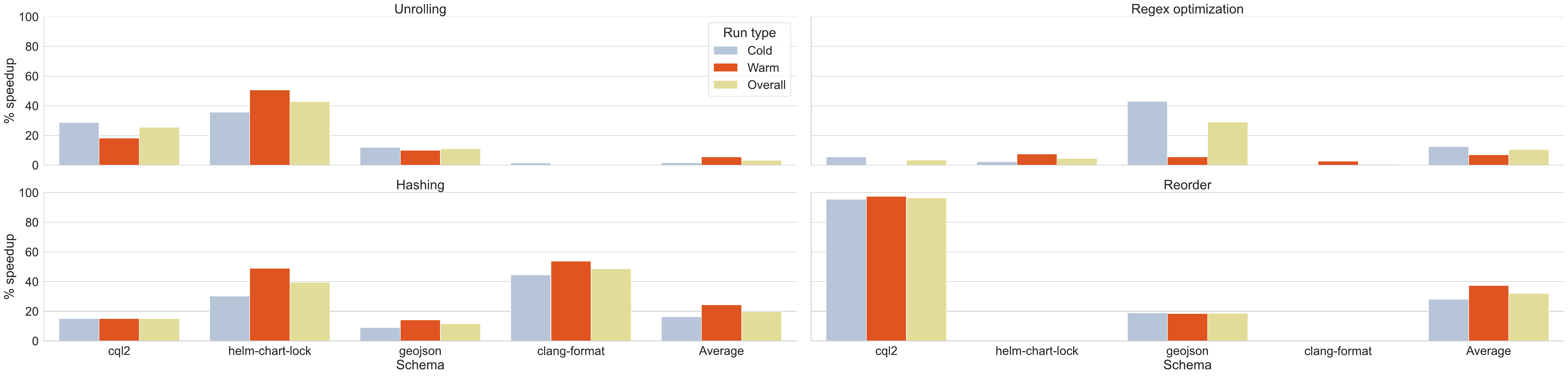}
    \caption{Performance benefits of various optimizations}\label{fig:ablation}
    \Description[Blaze optimizations]{A graph displaying performance impact of various optimizations for Blaze across different schemas}
\end{figure*}

\section{Related Work}\label{sec:related}

We are not aware of any in-depth academic research into JSON Schema validation beyond the work of Attouche et al.~\cite{modernjson}.
However, there are many existing open-source validators that make use of precompilation as we have listed in Table~\ref{tab:implementations}.
ajv\footnote{\url{https://ajv.js.org}} is a popular validator that is able to generate JavaScript code through a precompilation process to use for later validation.
While ajv is sometimes comparable to Blaze in speed, it suffers from significant correctness issues and fails over 200 test cases in the official JSON Schema test suite.
We believe that the speed of ajv comes primarily from the significant optimizations present in modern JavaScript runtimes.
Indeed, we noticed when switching from the Node.js runtime to the Bun runtime used in our evaluation, the validation performance improved significantly.
This observation matches what has been observed in prior work~\cite{ahmod2023javascript,kniazev2023choosing}.
Unlike ajv, Blaze maintains correctness in the validation process while also being faster to validate.

Corvus\footnote{\url{https://github.com/corvus-dotnet/Corvus.JsonSchema}}, jsonschema (Go)\footnote{\url{https://github.com/santhosh-tekuri/jsonschema}}, and JsonSchema.Net~\footnote{\url{https://docs.json-everything.net/schema/basics/}} also perform precompilation.
Corvus generates .NET code that compiled to produce the final validator while jsonschema constructs an internal data structure intended to optimize validation.
While these implementations are among the fastest for warm runs, they do not incorporate many optimizations used in Blaze and remain significantly slower to validate on average.
JsonSchema.Net performs static analysis of JSON Schema ~\cite{staticanalysis,jsonschemadotnet} in order to reduce the amount of work necessary at validation time by identifying constraints that must apply to particular locations in a JSON document.
Blaze extends this idea to significantly more in-depth analysis involving interactions between keywords and the order of instructions.

We also note that there has been some past work at examining the structure of JSON Schemas in practice~\cite{jsonschemas,defects,baazizi2021empirical}.
This work focused on four specific research questions that were not applicable to our analysis.
However, our overall methodology was similar.
We used the Sourcegraph public code search API\footnote{\url{https://sourcegraph.com/search}} to find files with extension \texttt{.json} and containing a key \texttt{\$schema} key indicating the document is a JSON Schema document.
We downloaded all these schemas and validated them against their corresponding meta-schema in order to ensure each schema is valid.
As mentioned previously, we collected approximately 31,000 schemas in total.
This allowed us to answer such questions as ``What is the distribution of key lengths defined in JSON Schemas?''.
We believe this corpus of schemas will be useful for further analysis.

\section{Future Work}\label{sec:future}

Currently, the schema compilation process results in a set of instructions that can be interpreted at runtime in a manner much more efficient than operating using the original JSON Schema.
However, we plan to explore the possibility of precompiling the code necessary to validate each schema ahead of time.
In addition to eliminating overhead from interpretation at runtime, this has the potential to leverage existing compiler optimizations to further improve performance.
We also plan to explore further static optimizations to the generated schemas.

We believe there is potential to optimize schema compilation in a data-dependent way.
Many instructions used for validation can be reordered while preserving correctness as we showed in Section~\ref{subsec:reordering}.
The fastest approach to validation will detect failure as early as possible.
Depending on the specific schema and the data being processed, it is possible that different use cases might result in a higher likelihood of certain assertions failing as compared to others.
If profiling suggests that a particular property is likely to fail validation, we can order instructions to validate that property first.
This enables early detection of validation failure, minimizing the number of executed instructions.
We also plan to explore the use of the \texttt{examples} keyword defined in JSON Schemas to drive data-dependent optimizations.

Finally, in this work we have focused only on indicating whether a document is valid according to a schema.
In the case of an invalid document, it can be helpful to provide information on exactly why the document is not accepted according to the schema.
This is particularly important in our case since we want to reference the user-provided schema ignoring any optimizations to instructions generated during the compilation process.
While Blaze does have the option to provide helpful error messages to users for debugging purposes, here we focus purely on performance.

\section{Conclusion}\label{sec:conclusion}

We have introduced Blaze, a JSON Schema validator that makes use of precompilation to optimize the validation process.
Unlike many existing validators, Blaze achieves 100\% correct validation behavior according to the JSON Schema specification.
Blaze also validates documents a minimum of 20\% faster than all other validators we tested on a wide variety of datasets and an average of 10$\times$ faster than the next fastest validator.
We believe that there are many opportunities for further optimization.

\bibliographystyle{ACM-Reference-Format}
\bibliography{references}

\end{document}